\documentclass[12pt]{iopart}
\usepackage{amsfonts}
\usepackage{graphicx}
\usepackage{amssymb}
\begin{document}
\title[Magnetic phases]{Magnetic phases of orbital bipartite optical lattices}

\author{Pil Saugmann and Jonas Larson}

\address{Department of Physics, Stockholm University, AlbaNova
  University Center, Se-106 91 Stockholm, Sweden}
\ead{pilmaria.saugmann@fysik.su.se}

\begin{abstract}
In the Hamburg cold atom experiment with orbital states in an optical lattice, $s$- and $p$-orbital atomic states hybridize between neighbouring sites. In this work we show how this alternation of sites hosting $s$- and $p$-orbital states gives rise to a plethora of different magnetic phases, quantum and classical. We focus on phases whose properties derive from frustration originating from a competition between nearest and next nearest neighbouring exchange interactions. The physics of the Mott insulating phase with unit filling is described by an effective spin-1/2 Hamiltonian showing great similarities with the $J_1$-$J_2$ model. Based on the knowledge of the $J_1$-$J_2$ model, together with numerical simulations, we discuss the possibility of realising a quantum spin liquid phase in the present optical lattice system. In the superfluid regime we consider the parameter regime where the $s$-orbital states can be adiabatically eliminated to give an effective model for the $p$-orbital atoms. At the mean-field level we derive a generalized classical $XY$ model, and show that it may support maximum frustration. When quantum fluctuations can be disregarded, the ground state is expected to be a spin glass. Even with quantum fluctuations present it has been debated whether a spin liquid may persist at the point of full frustration.
\end{abstract}
\pacs{1315, 9440T}
\submitto{\NJP}

\normalsize

\section{Introduction}
The physics of ultracold atoms has within the last two decades matured from laser cooling, trapping and precision spectroscopy to explorations of strongly interacting quantum many-body systems~\cite{mb}. With the aid of optical lattices and Feshbach resonances, the experimenter can emulate a variety of lattice models and carry out {\it in situ} measurements. The versatility of these systems in terms of altering lattice geometries, varying parameter strengths and initialize desired states, together with the possibility to perform high fidelity state measurements, make them perfect candidates for quantum simulations~\cite{qs}. In other words, utilize real physical systems in order to solve quantum mechanical problems that are intractable on classical computers. As realizations of lattice systems, prime candidates to be simulated are found in condensed matter physics such as Hubbard models~\cite{hubbard}. Following the theoretical proposal~\cite{zoller1}, the first experimental demonstration of the strongly correlated regime was the realization of the Bose-Hubbard model by Greiner {\it et al.}~\cite{bloch1}. This started an avalanche of activity, {\it e.g.} shortly afterwards it was suggested that internal atomic hyperfine levels could be utilised in order to mimic models of quantum magnetism~\cite{lukin}. The idea relies on the {\it superexchange mechanism}, where the effective spin-spin coupling derive from perturbation in terms of the tunneling terms. The first observation of superexchange interaction with cold atoms loaded in optical lattices was presented in~\cite{supex}. While it was shown that the spin-spin coupling could be tuned between supporting ferro- and anti-ferromagnetic order, the experiment could not explore different zero temperature magnetic phases due to too high effective temperatures. Instead, it took another couple of years before the first experiment could study such different magnetic phases~\cite{greiner}, and thereby also the corresponding quantum phase transitions (PTs). This idea did not make use of internal hyperfine states, but encoded the spin in occupations of the different sites. 

One group of phases that are of special interest are spin liquids or glasses. Such phases can emerge from strong {\it frustration}, which for their classical counterparts typically implies a large number of classically degenerate ground states~\cite{frustref}. Despite large fluctuations, order may build-up and at sufficiently low temperatures {\it quantum spin liquids} (QSL)~\cite{spinliq} can be formed. These phases are expected to host many novel properties like topological fractional excitations, which may find application in future quantum computing technologies~\cite{spinliqQC}. For solids, to date no direct evidences of quantum spin liquids have been demonstrated, but only indirect measurements that suggest their presence~\cite{spinliq}. When it comes to cold atoms in optical lattices, by using ``lattice shaking'' techniques the tunneling terms can be adjusted almost at will, and as a consequence magnetic frustration has been observed in these systems~\cite{sengstockexp}. While the experiment could only demonstrate classical frustration, a theoretical analysis predicts the existence of quantum spin liquid states in the same type of system set-up~\cite{sengstocktheo}.    

Orbital physics, {\it i.e.} the phenomena arising from the additional onsite degrees-of-freedom beyond charge and spin, plays an essential role in condensed matter physics. For example, for $p$-wave Cooper pairing of $^3$He~\cite{cooper} and for transition metal oxides which are crucial for high-$T_c$ superconductivity~\cite{htc}. Due to the importance in other branches, it did not take long until orbital physics was discussed in terms of cold atoms in optical lattices~\cite{girvin,xx}. It was also suggested that these orbital systems could be utilized to emulate magnetic models~\cite{fernanda}. The idea is to encode the synthetic spin degrees-of-freedom in orbital atomic states, rather than in internal hyperfine levels of the atoms. Whether orbital or internal hyperfine atom states are utilized result in qualitatively different effective spin models. For example, the structure of the different orbitals typically come about as anisotropic coupling terms. This is in contrast to effective spin degrees-of-freedom deriving from hyperfine states~\cite{lukin}, which normally do not render such anisotropy. Apart from the anisotropy, for bosonic atoms there is also an interaction term mixing different orbital states which is absent in systems where the spin is encoded in internal hyperfine states~\cite{fernanda}. Such a term typically breaks down a continuous symmetry into a discrete $\mathbb{Z}_n$-symmetry. 

In the pioneering work~\cite{girvin}, simulation of a $p$-orbital Bose-Hubbard model on a square optical lattice was considered. It was found that the superfluid (SF) state arrange in a vortex lattice. Such a state breaks time-reversal symmetry and can only be ascribed a complex order parameter. Furthermore, it was argued that the life-time of these excited states could be relatively long, {\it i.e.} relaxation into an ordered phase was expected, which was later confirmed experimentally~\cite{blochp}. For bosonic atoms, the main loss mechanism consists in scattering of two $p$-orbital atoms into one $s$- and one $d$-orbital atom. Even though the onsite optical lattice potential is not perfectly harmonic, the mentioned loss process is quasi-resonant and may become a serious hindrance when exploring $p$-orbital physics. To circumvent this, superlattices can be employed where such scattering processes are far off-resonant and thereby greatly suppressed. By hybridising degenerate $s$- and $p$-orbital atoms on neighbouring sites in a superlattice, this idea was experimentally verified in the Hamburg group of Hemmerich~\cite{hemmerich1,hemmerich2}. Also for this configuration, the superfluid phase consists of vortices leading to a complex order parameter~\cite{wu1}. It was further argued that thermal fluctuations will induce a new phase, a {\it chiral Bose liquid}, in this model~\cite{liu1}. A relevant question is if other novel phases may be realized with the Hemmerich set-up, and in particular if liquid or glass phases are possible. 

As we will show in this paper, much of the phase diagram of the Hemmerich set-up is unexplored. The physics of the insulating phases is, for example, not resolved. For the regular square lattice it is known that the first insulating Mott with one atom per site is effectively described by a spin-1/2 $XY\!Z$ model~\cite{fernanda} with a rich phase diagram already in one dimension. In the $s$-$p$ hybridised lattice the physics of this insulating phase is conceptually different; the $s$-orbital atoms play no direct role as they do not constitute any extra degree-of-freedom, but instead they serve as mediating the coupling between different sites with $p$-orbital atoms. Thus, the effective model consists solely of $p$-orbital atoms, and in particular, the strength of the coupling between nearest neighbours is of the same order as the coupling strength between next-nearest neighbours. Strong coupling to the next nearest neighbours is a possible route to achieve strong frustration~\cite{spinliq}. This gives hope to realize liquid phases. Indeed, our model resemblance the $J_1$-$J_2$ model that has served as a work horse in the study of QSL. Using mean-field methods and exact diagonalisation we see evidence for a new phase that agrees with earlier predictions of a QSL in the $J_1$-$J_2$ model. 

In the superfluid phase, the $s$-orbital atoms cannot be simply eliminated with the same argument as for the Mott since the onsite particle will typically fluctuate. However, if the $s$- and $p$-orbitals are far detuned in energy we may adiabatically eliminate the $s$-atoms, to again derive an effective model for the $p$-atoms. Just like for the insulating phase, the resulting model is inevitably constructed from nearest and next-nearest neighbour interactions. The sign of the detuning between the different orbitals determines also the sign of the tunneling strengths, which turns out to be desired in order to achieve frustration. For negative tunneling amplitudes the superfluid arranges in a phase with alternating vortices and anti-vortices between neighbouring sites, {\it i.e.} breaking time-reversal symmetry and with a complex order parameter. For positive tunneling amplitudes the system shows frustration. In particular, at the mean-field level we derive a generalization of a classical $XY$ model. Like for the $J_1$-$J_2$ model, the $XY$ model has also served as a prototype in order to explore frustration and spin liquid phases. We find support for a frustration-driven phase also in our model which presumably is a spin liquid.  
 
The reminder of the paper is structured as follows. In the next section we introduce the physical system of hybridised $s$-$p$ orbitals in an optical lattice.  We derive the full Hamiltonian in the tight-binding limit, from which we derive the effective models used to describe the physics of the insulating phase and superfluid phase in subsec. 2.2 and 2.3 respectively. The proceeding sec. 3 analyses the phase diagrams of the respective models, and we discuss the prospects of phases driven by frustration. Finally we conclude with a summary in sec. 4.

\section{Model Hamiltonian}
\subsection{Physical system and its Hamiltonian}
\begin{figure}[h]
\centerline{\includegraphics[width=10cm]{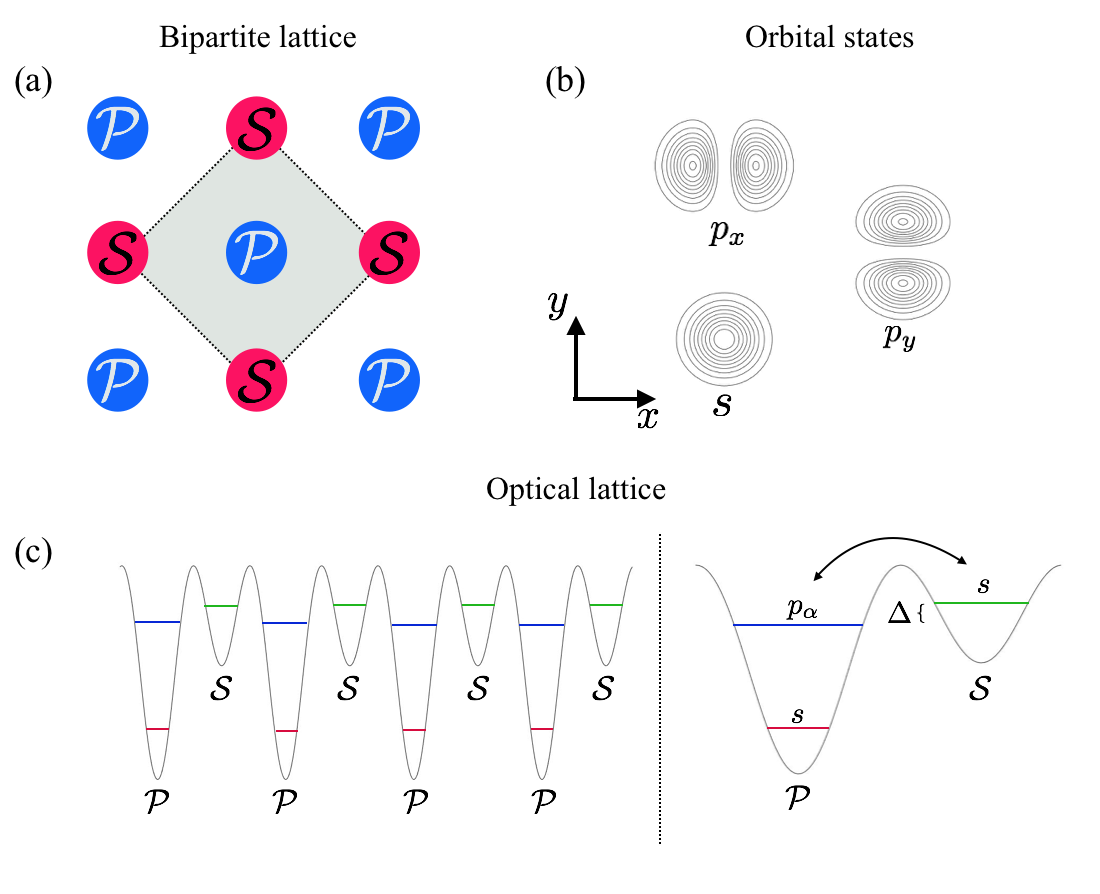}}
\caption{Structure of the optical lattice potential. Visualized in (a), the superlattice produces alternating deep ($\mathcal{P}$) and shallow ($\mathcal{S}$) potential wells. The unit cell is marked by the grey square and comprises one $\mathcal{S}$- and one $\mathcal{P}$-site. The relevant orbital states are shown in (b); a $p_x$-orbital has a node in the $x$-direction, while a $p_y$-orbital has a node in the $y$-direction. Thus, the $p$-orbitals change sign across the node, positive on one side and negative on the other. The $s$-orbital has no node and is polar symmetric. A cut of the two dimensional lattice is pictured in (c) showing the alternating $\mathcal{S}$- and $\mathcal{P}$-sites. The horizontal lines represent the onsite energies of the different orbitals; red -- $\mathcal{P}$-site $s$-orbital, blue -- 
$\mathcal{P}$-site $p$-orbital and green -- $\mathcal{S}$-site $s$-orbital. By construction, three orbital states are quasi resonant; $p_x$ and $p_y$ on the $\mathcal{P}$-sites and $s$ on the $\mathcal{S}$-sites. The parameter $\Delta$ sets the energy difference between the average of the two $p$-orbital states and the $s$-orbital states. Not shown in the figure is the energy splitting between the $p_x$- and $p_y$-orbitals. } \label{fig1}
\end{figure}

The optical potential of the Hemmerich experiment~\cite{hemmerich1,hemmerich2} forms two sub-lattices, $\mathcal{S}$ and $\mathcal{P}$ with the $\mathcal{P}$-sites deeper, see Fig~\ref{fig1} (a). By tuning the lattice parameters, the relative depth of the $\mathcal{S}$- and $\mathcal{P}$-sites is chosen such that the two $p_x$- and $p_y$-orbitals on the $\mathcal{P}$-sites are quasi resonant with the $s$-orbitals on the $\mathcal{S}$-sites (Fig.~\ref{fig1} (c)). The atoms are assumed to only populate the corresponding three bands made out of the two $p$- and the $s$-orbital states. We neglect any losses to other bands. Furthermore, as is usually assumed, we consider the {\it tight-binding approximation}, consisting in only taking nearest neighbour tunneling and onsite $s$-wave scattering into account.

The $p_x$- and $p_y$-orbital states have, respectively, a node in the $x$- and $y$-direction as shown in Fig.~\ref{fig1} (b), while the $s$-orbitals are polar symmetric. In the general case, the extended width of the $p$-orbitals in either direction causes anisotropic tunnelings -- a $p_x$-orbital atom is more likely to tunnel in the $x$-direction (also called $\pi$-bonding in orbital physics) than in the $y$-direction (or $\sigma$-bonding), and vice versa for a $p_y$-orbital atom. In the present bipartite lattice, however, where the tunneling occurs between consecutive $\mathcal{S}$- and $\mathcal{P}$-sites, the tunneling in the $\alpha$-direction ($\alpha=x,\,y$) is possible only for a $p_\alpha$ orbital atom, {\it i.e.} $p_x$-orbital atoms tunnel only in the $x$-direction and correspondingly for the $p_y$-orbital atoms. This follows from the parity of the orbital states together with the shape of the optical potential. There are two ways for the $p_x$- and $p_y$-orbital atoms to couple; a $p_x$-orbital atom can tunnel to an $s$-orbital atom which then tunnels to a $p_y$-orbital atom, or two $p_x$-orbital atoms can scatter into two $p_y$-orbital atoms or vice versa. The latter process, the scattering involving different atomic states, is approximately zero for most spinor condensates. That is, the amplitude for scattering two atoms in one specific hyperfine state into two atoms in another hyperfine state is very small, typically one or two percents of any other scattering processes, while for orbital states all scattering processes are of the same order. It is clear that these processes break the particle number conservation of different atomic species (atoms in different orbital states, or in different hyperfine states). 

We denote the corresponding bosonic annihilation operators $\hat a_{\alpha\mathbf{i}}$ ($\alpha=x,\,y$) for the $p_\alpha$-orbitals and $\hat a_{s\mathbf{i}}$ for the $s$-orbitals (and similarly for the creation operators). The subscript $\mathbf{i}=(i_x,i_y)$ gives the site, i.e. $i_\alpha\in Z$. When restricting the model to these three bands, an atomic operator annihilating an atom at ${\bf x}=(x,y)$ is expanded as
\begin{equation}
\hat\Psi(x)=\sum_{\mathbf{i}\in\mathcal{S}}\hat a_{s\mathbf{i}}w_{s\mathbf{i}}({\bf x})+\sum_\alpha\sum_{\mathbf{j}\in\mathcal{P}}\hat a_{\alpha\mathbf{j}}w_{\alpha\mathbf{j}}({\bf x}),
\end{equation}
where $w_{s\mathbf{i}}({\bf x})$ and $w_{\alpha\mathbf{j}}({\bf x})$ are the corresponding Wannier functions localized at site $\mathbf{i}$ and $\mathbf{j}$, and their shapes reproduce the characteristics of the orbitals of Fig.~\ref{fig1} (b). The tunneling amplitude for a $p_x$-orbital atom into an $s$-orbital atom is called $t_x$, and similarly $t_y$ is the tunneling for a $p_y$-orbital atom. In the isotropic lattice, $t_x=t_y>0$. When giving up the isotropy of the lattice slightly this equality is not strictly true, and the two $p$-orbital states are no longer degenerate. We denote the onsite energy difference between the two $p$-orbitals by $\delta$ and let the zero energy lie exactly between these two. We can divide the full Hamiltonian $\hat H=\hat H_\mathrm{kin}+\hat H_\mathrm{int}$ into a kinetic (and onsite) term
\begin{equation}\label{ham1}
\begin{array}{lll}
\hat H_\mathrm{kin} & = & \displaystyle{-t_x\sum_{\langle\mathbf{ij}\rangle_x}\left(\hat a_{s\mathbf{i}}^\dagger\hat a_{x\mathbf{j}}+h.c.\!\right)}\\ \\
& & \displaystyle{\!-t_y\sum_{\langle\mathbf{ij}\rangle_y}\!\left(\hat a_{s\mathbf{i}}^\dagger\hat a_{y\mathbf{j}}+h.c.\!\right)\!+\Delta\!\sum_\mathbf{i}\hat n_{s\mathbf{i}}\!+\frac{\delta}{2}\sum_\mathbf{j}\left(\hat n_{x\mathbf{j}}-\hat n_{x\mathbf{j}}\right)},
\end{array}
\end{equation}
and the interaction term $\hat H_\mathrm{int}=\hat H_{nn}+\hat H_\mathrm{fc}$, which can be further decomposed into `density-density' interactions
\begin{equation}\label{ham2}
\hat H_{nn}\!=\!\sum_\alpha\!\sum_{\mathbf{j}\in \mathcal{P}}\!\!\frac{U_{\alpha\alpha}}{2}\hat n_{\alpha\mathbf{j}}\!\left(\hat n_{\alpha\mathbf{j}}\!-\!1\right)+\sum_{\mathbf{i}\in \mathcal{S}}\!\!\frac{U_{ss}}{2}\hat n_{s\mathbf{i}}\!\left(\hat n_{s\mathbf{i}}\!-\!1\right)+\!\!\sum_{\alpha\beta,\,\alpha\neq\beta}\!\sum_{\mathbf{j}\in \mathcal{P}}\!U_{\alpha\beta}\hat n_{\alpha\mathbf{j}}\hat n_{\beta\mathbf{j}}
\end{equation}
and `flavour-changing' interactions
\begin{equation}\label{ham3}
\hat H_\mathrm{fc}=\sum_{\alpha\beta,\,\alpha\neq\beta}\sum_{\mathbf{j}\in \mathcal{P}}\frac{U_{\alpha\beta}}{2}\left(\hat a_{\alpha\mathbf{j}}^\dagger\hat a_{\alpha\mathbf{j}}^\dagger\hat a_{\beta\mathbf{j}}\hat a_{\beta\mathbf{j}}+h.c.\right).
\end{equation}
Here, $\langle\dots\rangle_x$ and $\langle\dots\rangle_y$ indicate summing over nearest neighbours in the $x$- and $y$-direction respectively, and we have used the subscript $\mathbf{j}$ for $\mathcal{P}$-sites and $\mathbf{i}$ for $\mathcal{S}$-sites. The operators $\hat n_{\alpha\mathbf{j}}=\hat a_{\alpha\mathbf{j}}^\dagger\hat a_{\alpha\mathbf{j}}$ and $\hat n_{s\mathbf{i}}=\hat a_{s\mathbf{i}}^\dagger\hat a_{s\mathbf{i}}$ give the number of orbital atoms on the specific site. With the $p_x$- and $p_y$-orbitals having an onsite energy $\delta/2$ and $-\delta/2$ respectivelyt, while the onsite energy for the $s$-orbitals is $\Delta$. The interaction terms alone support a $\mathbb{Z}_2$-parity symmetry that represents conservation of orbital atoms modulo 2 -- the flavour-changing interaction converts two atoms of one orbital type into two atoms of another orbital type. This symmetry is, however, broken when the tunneling is allowed since $p$-orbital atoms can tunnel into $s$-orbital atoms, and thereby changing the total number of $p$-orbital atoms. In the isotropic lattice and when $\delta=0$, there is $\mathbb{Z}_4$ symmetry, corresponding to 90 degree rotations, that survives for the full many-body Hamiltonian. This corresponds to swapping the $p_x$- and $p_y$-orbitals and rotate the axes, {\it i.e.} the subscripts in the Hamiltonian transforms as $x\leftrightarrow y$ and $\mathbf{j}=(j_x,j_y)\rightarrow\mathbf{j}'=(-j_y,j_x)$. When the lattice is anisotropic the $\mathbb{Z}_4$ symmetry breaks down into a $\mathbb{Z}_2$ represented by a 180 degree rotation, $\mathbf{j}\rightarrow\mathbf{j}'=-\mathbf{j}$. The total particle number is, of course, another preserved quantity, and it is this continuous $U(1)$ symmetry that is broken across the {\it Mott-superfluid} PT~\cite{jani2}. The onsite energies $\delta$ and $\Delta$, the relative interaction strengths $U_{\alpha\beta}$ and the tunneling amplitudes $t_\alpha$ are all determined by the overlap integrals of the Wannier functions. For example, the interaction strengths are
\begin{equation}
U_{\alpha\beta}=U_0\int d{\bf x}\,|w_{\alpha\mathbf{j}}({\bf x})|^2|w_{\beta\mathbf{j}}({\bf x})|^2,\hspace{0.5cm}U_{ss}=U_0\int d{\bf x}\,|w_{s\mathbf{j}}({\bf x})|^4,
\end{equation}
with $U_0$ a constant proportional to the $s$-wave scattering length. In the harmonic approximation, the Wannier functions are replaced with harmonic functions, {\it e.g.} (in dimensionless units)
\begin{equation}
w_{x\mathbf{j}}({\bf x})=\left(\frac{2}{\pi\sigma^4}\right)^{1/2}(x-x_{j_x})\exp\left(-\frac{(x-x_{j_x})^2}{2\sigma^2}-\frac{(y-y_{j_y})^2}{2\sigma^2}\right),
\end{equation}
where $(x_{j_x},y_{j_y})$ is the position of site $\mathbf{j}=(j_x,j_y)$, and we have assumed the width $\sigma$ to be the same in the $x$- and $y$-directions. The interaction strengths then obey $U_{xx}=U_{yy}=3U_{xy}$~\cite{jani}. The $U_{ss}$ is not simply related to the other interaction parameters since it derives from the interaction in an $\mathcal{S}$- and not a $\mathcal{P}$-site, but it is of the same order as $U_{\alpha\alpha}$. 

When considering a single site on the $\mathcal{P}$-sub-lattice it is convenient to introduce the angular momentum operator
\begin{equation}\label{angop}
\hat L_{z\mathbf{j}}=-i\left(\hat a_{x\mathbf{j}}^\dagger\hat a_{y\mathbf{j}}-\hat a_{y\mathbf{j}}^\dagger\hat a_{x\mathbf{j}}\right)
\end{equation}
and the total number operator $\hat n_\mathbf{j}=\hat n_{x\mathbf{j}}+\hat n_{y\mathbf{j}}$. Then, up to a term proportional to $\hat n_\mathbf{j}$, the interaction terms can be written
\begin{equation}
\hat H_\mathrm{int,\mathbf{j}}^{(\mathcal{P})}=\frac{U_0}{2}\left(\hat n_\mathbf{j}^2-\frac{1}{3}\hat L_{z\mathbf{j}}^2\right).
\end{equation}
Thus, for repulsive interaction $U_0>0$, the interaction energy is minimized by maximizing the $z$ angular momentum $\hat L_{z\mathbf{j}}$. In the superfluid phase, the onsite order parameter should then be chosen $\psi_\mathbf{j}({\bf x})=w_{x\mathbf{j}}(\mathbf{x})\pm iw_{y\mathbf{j}}({\bf x})$, which represents a clockwise or anti-clockwise vortex state with $\pm1$ vorticity~\cite{girvin,liuwu,wufrust1}.  

\subsection{Large detuning effective model}
When the energy off-set $\Delta$ between $p$- and $s$-orbital states is small, the two types of atomic states hybridize to build up the full system state~\cite{jani2}. On the other hand, when $|\Delta|$ is the large parameter (compared to the tunneling terms and to $U_{ss}$), population transfer between the $\mathcal{S}$- and the $\mathcal{P}$-sites is hindered due to the large energy difference. The kinetics is still not trivial, {\it i.e.} frozen, as the intermediate states can virtually mediate tunneling between next-nearest neighbors in a two-step process. 

\begin{figure}[h]
\centerline{\includegraphics[width=4cm]{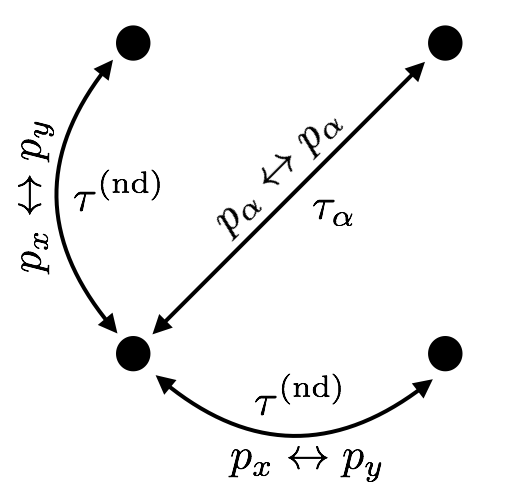}}
\caption{Tunneling terms between nearest and next nearest neighbouring sites (black dots) in the effective Hamiltonian~(\ref{detham}). Between nearest neighbours tunneling is accompanied with a flip in the orbital state reminiscent of a spin-orbit coupling. Along the diagonal, next nearest neighbours, the orbital state is not changed upon tunneling, and $p_x-p_x$ couplings are along one diagonal and $p_y-p_y$ along the other. The respective amplitudes for the two processes are denoted $\tau^{\mathrm{(nd)}}$ and $\tau_\alpha$. } \label{fig2}
\end{figure} 

In the~\ref{appa} we give the derivation of the effective model that results from the elimination of the $s$-orbitals. In the general case with non-zero transverse and longitudinal tunnelings the number of terms becomes very large. Limiting the analysis to only longitudinal tunneling reduces the number considerably, and this is also the case relevant for the Hemmerich experimental set-up where the transverse tunneling vanishes~\cite{hemmerich1,hemmerich2}. The resulting Hamiltonian reads 
\begin{equation}\label{detham}
\begin{array}{lll}
\hat H_\mathrm{eff}^{(\Delta)} & = & \displaystyle{\!\!-\sum_\alpha\sum_{\{\mathbf{ij}\}_\alpha}\tau_\alpha\hat a_{\alpha\mathbf{i}}^\dagger\hat a_{\alpha\mathbf{j}}-\!\!\sum_{\alpha\beta,\,\alpha\neq\beta}\sum_{\langle\mathbf{ij}\rangle}\left(\tau^{\mathrm{(nd)}}\hat a_{\alpha\mathbf{i}}^\dagger\hat a_{\beta\mathbf{j}}+h.c.\right)}\\ \\
& & \displaystyle{\!\!-\sum_\alpha\sum_{\mathbf{j}}\tau_\alpha\hat n_{\alpha\mathbf{j}}+\frac{\delta}{2}\sum_\mathbf{j}\!\left(\hat n_{x\mathbf{j}}-\hat n_{y\mathbf{j}}\right)\!+\!\sum_\alpha\sum_{\mathbf{j}}\frac{U_{\alpha\alpha}}{2}\hat n_{\alpha\mathbf{j}}\!\left(\hat n_{\alpha\mathbf{j}}-1\right)\!}\\ \\
& & +\displaystyle{\sum_{\alpha\beta,\,\alpha\neq\beta}\sum_{\mathbf{j}}U_{\alpha\beta}\hat n_{\alpha\mathbf{j}}\hat n_{\beta\mathbf{j}}+\sum_{\alpha\beta,\,\alpha\neq\beta}\sum_{\mathbf{j}}\frac{U_{\alpha\beta}}{2}\left(\hat a_{\alpha\mathbf{j}}^\dagger\hat a_{\alpha\mathbf{j}}^\dagger\hat a_{\beta\mathbf{j}}\hat a_{\beta\mathbf{j}}+h.c.\right)},
\end{array}
\end{equation}
where $\tau_\alpha=|t_\alpha|^2/\Delta$ and $\tau^\mathrm{(nd)}=t_xt_y/\Delta$ give the tunneling amplitudes along nearest and next nearest neighbours respectively, see Fig.~\ref{fig2}. Thus, the curly bracket $\{\mathbf{ij}\}_\alpha$ in the sum denotes summing over next nearest neighbours in the direction $\alpha$, and as before $\langle\mathbf{ij}\rangle$ sums instead over nearest neighbours. Note that the two tunneling strengths are of the same order, {\it e.g.} if we consider an isotropic lattice, $t_x=t_y$, we have  $\tau_x=\tau_y=\tau^\mathrm{(nd)}\equiv\tau$. What is particularly appealing is that the signs of the tunneling amplitudes is adjustable by changing the sign of $\Delta$ ({\it i.e.} considering red or blue detuning). Tunneling between nearest neighbours induces a swapping of the orbital state, which can be seen as an effective spin-orbit coupling -- the `spin' (orbital) degree-of-freedom is coupled to the motion of the atom~\cite{wuso}. The third sum derives from `Stark shifts' due to the virtual couplings to the $s$-orbitals. 

\subsection{Large interaction effective model}\label{sec2.3}
The models described above are good starting points for exploring the superfluid phase. For the insulating phases it is convenient to expand in powers of $t/U$, {\it i.e.} the tunneling amplitude over the interaction strength~\cite{auerbach}. By further projecting the model to a fixed number of atoms per site the Hilbert space dimension is greatly reduced and the physics becomes easier to analyse. We will only consider the {\it Mott}$_1$ insulating phase with unit filling as this is the experimentally most relevant one~\cite{hemcom}, and already at this level one finds an interesting effective model Hamiltonian. The details of the derivation can be found in the~\ref{appb}. The method is rather standard, and has also been applied to $p$-orbital atoms loaded in optical lattices~\cite{fernanda}. However, there are a few ingredients that are special for our system that are worth pointing out. 

The flavour-changing interaction term~(\ref{ham3}) is clearly non-diagonal in the Fock basis which causes the allowed processes in the perturbation expansion to be  much richer; the interaction does not only add a constant which is the case for the density-density interaction terms. This implies that application of a simple {\it hard core boson} approach gives qualitatively wrong effective models. On the separable square lattice of $p_x$- and $p_y$-orbitals, where a given orbital atom can tunnel both along the direction of its node and in its perpendicular direction, non-trivial contributions appear already at second order~\cite{fernanda}. For the bipartite lattice of alternating $p$- and $s$-orbital sites, the $p$-orbital atoms are only allowed to tunnel in the direction of its node. The second-order terms are then trivial. As an example, take a $p_x$-orbital atom at a given site, it can tunnel to its neighbouring $s$-site with amplitude $t_x$ such that there are two $s$-orbital atoms at that site giving an interaction contribution $\sim1/U_{ss}$, and then one of the $s$-orbital atoms tunnel back to the empty $p$-site as a $p_x$-orbital atom and again with an amplitude $t_x$. One gets a term $\sim\left(|t_x|^2/U_{ss}\right)\hat a_{x\mathbf{j}}^\dagger\hat a_{x\mathbf{j}}$, which is nothing but an energy shift. If the lattice is isotropic, both the $p_x$- and $p_y$-orbitals pick up the same energy shift. Thus, via second-order processes it is not possible to generate effective interaction terms between the orbitals, so to reach a non trivial effective Hamiltonian one must include fourth order terms as well. This gives rise to nearest and next nearest neighbour couplings for the $\mathcal{P}$-sites. Note that since there is no internal degree-of-freedom in the $\mathcal{S}$-sites, these will effectively freeze out. There are essentially two different types of tunneling processes, non-loops and loops, in the non-loop process the atom tunnels out and back along the same path which is not the case in the loop processes. The next nearest neighbouring couplings consist only of non-loop processes, while the nearest neighbouring couplings have both non-loop and loop contributions. The non-loop processes give rise to a density-density coupling, while the loop processes result in orbital swapping couplings. The couplings of nearest neighbouring $\mathcal{P}$-sites consist of both non-loop contribution on the form $\sim\left(|t|^4/(U_{ss}^{2}U_{pp})\right)\hat n_{x\mathbf{i}}^\dagger\hat n_{y\mathbf{j}}$ and loop contributions on the form $\sim\left(|t|^4/(U_{ss}^{2}U_{pp})\right)\hat a_{x\mathbf{i}}^\dagger\hat a_{y\mathbf{i}}\hat a_{x\mathbf{j}}^\dagger\hat a_{y\mathbf{j}}$, while the next nearest neighbouring couplings only consist of non-loop couplings as  $\sim\left(|t|^4/(U_{ss}^{2}U_{pp})\right)\hat n_{\alpha\mathbf{i}}^\dagger\hat n_{\alpha\mathbf{j}}$. Here $|t|^4$ is the corresponding four tunneling amplitudes ({\it i.e.} $|t_\alpha|^4$ or $|t_x|^2|t_y|^2$) and $U_{pp}$ the same for the interactions. 

With the restriction of single particle occupancy on every site it is practical to map the Hamiltonian into one of spin-$1/2$ particles by using the $p$-orbital states to define the {\it Schwinger spin bosons}~\cite{auerbach}
\begin{equation}\label{Scwinger}
\begin{array}{lll}
 \hat{S}_\mathbf{i}^{Z}&=&\frac{1}{2}\left(\hat{a}_{x_\mathbf{i}}^{\dagger}\hat{a}_{x_\mathbf{i}}-\hat{a}_{y_\mathbf{i}}^{\dagger}\hat{a}_{y_\mathbf{i}}\right),\\ \\
 \hat{S}_\mathbf{i}^{+}&=&\hat{S}_\mathbf{i}^{X}+i\hat{S}_\mathbf{i}^{Y}= \hat{a}_{x_\mathbf{i}}^{\dagger}\hat{a}_{y_\mathbf{i}},\\ \\
 \hat{S}_\mathbf{i}^{-}&=&\hat{S}_\mathbf{i}^{X}-i\hat{S}_\mathbf{i}^{Y}= \hat{a}_{y_\mathbf{i}}^{\dagger}\hat{a}_{x_\mathbf{i}}.
\end{array}
\end{equation}
With this mapping the effective Hamiltonian takes the form
\begin{equation}
\begin{array}{lll}\label{HamMott}
\hat H_\mathrm{eff}& = &\displaystyle{h^Z\sum_\mathbf{i}\hat S_\mathbf{i}^Z+J_2^{Z}\sum_{\left\{\mathbf{ij}\right\}}\hat S_\mathbf{i}^{Z}\hat S_\mathbf{j}^{Z}}\\ \\
& &\displaystyle{+\sum_{\langle\mathbf{ij}\rangle}\left(J_1^{X}\hat S_\mathbf{i}^{X}\hat S_\mathbf{j}^{X}+J_1^{Y}\hat S_{i}^{Y}\hat S_{j}^{Y}+J_1^Z\hat S_\mathbf{i}^{Z}\hat S_\mathbf{j}^{Z}\right)}.
\end{array}
\end{equation}
The first sum mimics a field in the $Z$-direction with an amplitude $h^Z$~\cite{fernanda0}, the subscripts on the coupling amplitudes indicates whether the sum is over nearest, 1, or next nearest neighbouring, 2, sites (as before, the curly brackets stand for summing over next nearest neighbours). The explicit expressions for the coupling amplitudes are presented in the~\ref{appb}, and here we only point out the relation $J_1^{Z}=-J_2^{Z}/2$ that is of importance when discussing the possible phases in the next section. Without the second term over next nearest neighbours the model comprises the two dimensional $XY\!Z$ Heisenberg model~\cite{xyz}.

\section{Phase diagrams}
Topological aspects of interacting fermionic atoms loaded in the same type of bipartite optical lattice as considered in this work have been discussed in the past~\cite{topo}. For bosonic atoms, relevant for us, the Mott insulator-superfluid phase boundaries were studied in~\cite{jani2} using the Gutzwiller mean-field method. Also the properties of the superfluid phase was analyzed in~\cite{jani2}, but the non-zero detuning $\Delta\neq0$ situation has so far been unnoticed. Likewise, the properties of the insulating phases are still to be explored. The phases of the Mott insulating phase with one particle per site was considered in Ref.~\cite{fernanda} for the separable square lattice of $p_x$- and $p_y$-orbitals. As already pointed out, the present system is conceptually different since $s$-orbitals induce effective couplings between sites beyond nearest neighbours. The goal is to determine whether the interplay between neighbouring couplings can give rise to novel phenomena like glassiness and spin-liquids~\cite{j1j2}.  

\subsection{Superfluid phase diagram}
\subsubsection{Zero detuning phase diagram}
As we pointed out, the $\Delta=0$ situation was already explored experimentally in~\cite{hemmerich1} and theoretically in~\cite{jani2}. Nevertheless, we revisit it here for completeness, and this will also help us in better understanding the non-zero detuning case in the subsequent section.

Deep in the SF phase quantum fluctuations play a less important role, and within the simplest mean-field approximation we assign a coherent state to every boson mode. Thus, the full state can be expressed as
\begin{equation}
|\Psi\rangle=\prod_{\mathbf{i}\in\mathcal{S}}|\alpha_{s\mathbf{i}}\rangle_\mathbf{i}\prod_{\mathbf{j}\in\mathcal{P}}|\alpha_{x\mathbf{j}},\alpha_{y\mathbf{j}}\rangle_\mathbf{j},
\end{equation}
where $\hat a_{s\mathbf{i}}|\alpha_{s\mathbf{i}}\rangle_\mathbf{i}=\alpha_{s\mathbf{i}}|\alpha_{s\mathbf{i}}\rangle_\mathbf{i}$ and equivalently for the $x$- and $y$-modes. It is understood that the full state is a direct product state between the different coherent states, such that we have neglected any quantum correlations between different modes. The full condensate order parameter becomes
\begin{equation}\label{sforder}
\Psi(x)=\sum_{\mathbf{i}\in\mathcal{S}}\alpha_{s\mathbf{i}}w_{s\mathbf{i}}({\bf x})+\sum_{\mathbf{j}\in\mathcal{P}}\bigg[\alpha_{x\mathbf{j}}w_{x\mathbf{j}}({\bf x})+\alpha_{y\mathbf{j}}w_{y\mathbf{j}}({\bf x})\bigg].
\end{equation}
The (complex) coherent state amplitudes are determined from minimizing the energy functional $E[\alpha_{s\mathbf{i}},\alpha_{x\mathbf{j}},\alpha_{y\mathbf{j}}]=\langle\Psi|\hat H|\Psi\rangle$. Provided that the Hamiltonian is normally ordered, the energy functional is obtained from replacing the boson operators with their respective coherent state amplitudes, {\it e.g.} $\hat a_{s\mathbf{i}}\rightarrow\alpha_{s\mathbf{i}}$ and $\hat a_{s\mathbf{i}}^\dagger\rightarrow\alpha_{s\mathbf{i}}^*$. The absolute value squared of the complex amplitudes give the onsite atom numbers, $n_{s\mathbf{i}}=|\alpha_{s\mathbf{i}}|^2$, $n_{x\mathbf{j}}=|\alpha_{x\mathbf{j}}|^2$ and $n_{y\mathbf{j}}=|\alpha_{y\mathbf{j}}|^2$, while the onsite phases $\phi_{s\mathbf{i}}$,  $\phi_{x\mathbf{j}}$ and  $\phi_{y\mathbf{j}}$ ({\it e.g.} $\alpha_{s\mathbf{i}}=\sqrt{n_{s\mathbf{i}}}\exp(i\phi_{s\mathbf{i}})$) determine the global condensate coherence.

\begin{figure}[h]
\centerline{\includegraphics[width=11cm]{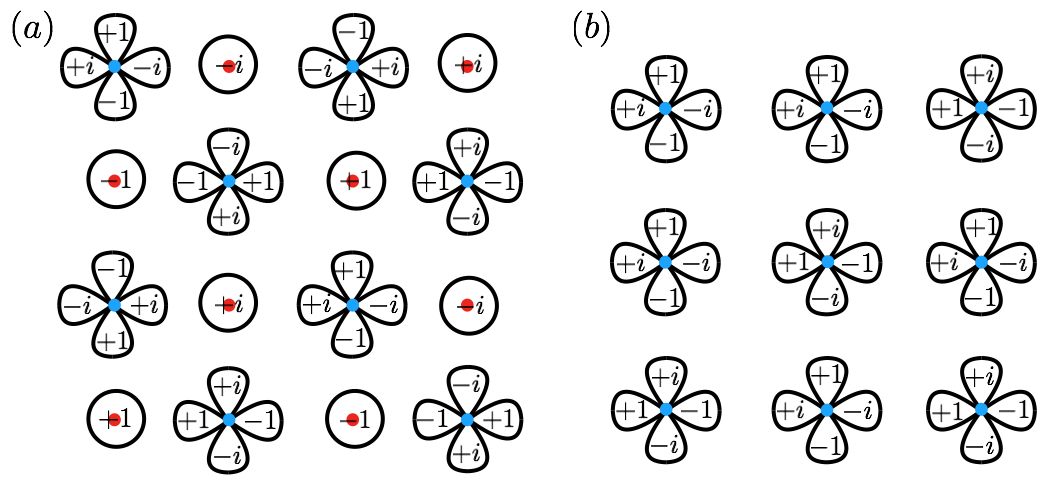}}
\caption{Global condensate phase order of the SF phase in the case of vanishing detunings $\Delta=\delta=0$ (a), and for large detuning $\Delta$ and with positive tunneling $\tau$ (b) (corresponding mean-field energy~(\ref{mfen})). The $\mathcal{P}$-lattice is denoted by the blue dots, and the $\mathcal{S}$-lattice by red dots. In both cases, at each $p$-site the condensate arranges itself in a vortex state which minimizes the interaction energy, {\it i.e.} the relative phase between the $p_x$- and $p_y$-orbitals is $\pi/2$, while the overall onsite phase is determined from minimizing the kinetic energy. When simultaneously minimizing the kinetic and interaction energies, the $p$-sites order in either vortices or anti-vortices on every site when $\Delta=0$ (a) which reflects a spontaneously broken $\mathbb{Z}_4$-symmetry~\cite{hemmerich1,jani2}, or in a checkerboard state of alternating vortices and anti-vortices (b) and a broken $\mathbb{Z}_2$-symmetry. } \label{fig3}
\end{figure} 

We have already pointed out in the previous section, that if $\delta=0$ the onsite interaction on the $\mathcal{P}$-sub-lattice is minimized by clockwise or anti-clockwise singly excited vortices, which are represented by a $\pi/2$ relative phase between $\phi_{x\mathbf{j}}$ and  $\phi_{y\mathbf{j}}$. This is easily seen by writing down the corresponding onsite energy functional for the interaction terms
\begin{equation}
E_\mathrm{int}^{(\mathbf{j})}[n_{x\mathbf{j}},n_{y\mathbf{j}},\phi_{x\mathbf{j}},\phi_{y\mathbf{j}}]\!=\!\frac{U_0}{2}\!\left(\!n_{x\mathbf{j}}^2\!+\!n_{y\mathbf{j}}^2\!+\!\frac{2}{3}n_{x\mathbf{j}}n_{y\mathbf{j}}\!\right)\!+\frac{U_0}{3}n_{x\mathbf{j}}n_{y\mathbf{j}}\cos(2(\phi_{x\mathbf{j}}-\phi_{y\mathbf{j}})).
\end{equation}
Since $U_0>0$, the minimum is obtained for $\phi_{x\mathbf{j}}-\phi_{y\mathbf{j}}=(2l+1)\pi/2$ for any integer $l$, and $n_{x\mathbf{j}}=n_{y\mathbf{j}}$. Given the vortex/anti-vortex solutions on the $p$ sites, the question is whether such solutions are also consistent with minimizing the kinetic energy, and thereby the total energy functional. 

For single $p$ sites, the overall phase is not affecting the onsite energy, but such phases become important when minimizing the total energy. These phases determine the global coherence of the condensate in the lattice. Already in Ref.~\cite{hemmerich1} it was noticed that it is possible to keep the vortices and still lock the onsite phases such that the kinetic energy is also minimized. For the isotropic case, $\delta=0$, the solution is pictured in Fig.~\ref{fig3}. It is seen that the $s$-orbitals lock the phases of the $p$-orbital vortices; each vortex in the lattice either circulate clockwise or anti-clockwise (they differ only in their overall phase). In the thermodynamic limit, the ground state is fourfold degenerate~\cite{jani2}, such that when the atoms condensate it spontaneously breaks a $\mathbb{Z}_4$-symmetry. It is also clear that time-reversal symmetry must be broken.   

When $\delta\neq0$, one of the two orbital states, $p_x$ or $p_y$, is energetically favoured. This split the fourfold degeneracy down to a doubly degenerate ground state, {\it i.e.} the system supports a $\mathbb{Z}_2$-symmetry. With the Schwinger spin operators introduced in (\ref{Scwinger}), at the mean-field level, an onsite vortex/anti-vortex state can be seen as an eigenstate of $\hat S_\mathbf{j}^y$, while the state with only the $p_x$- or $p_y$-mode populated is a corresponding eigenstate of $\hat S_\mathbf{j}^z$. Thus, the vortex solution found for $\delta=0$ can be viewed as an anti-ferromagnetic state in the $y$-direction, and when $\delta\neq0$ the system orders in a ferromagnetic (polarized) state in the $z$-direction. There is a first-order transition separating the two phases.

When $\delta\neq0$, the mean-field solution is still easy to find analytically. In particular, the sign of $\delta$ determines whether the state is polarised in the positive or negative $z$-direction. The two states are separated by a first order phase transition, and on the symmetry point $\delta=0$ the state forms the aforementioned anti-ferromagnetic state in the $y$-direction.

\subsubsection{Large detuning phase diagram}
We saw that for $\Delta=0$ it is possible to find an analytic mean-field solution that minimizes simultaneously the onsite interaction and the kinetic energies. Thus, the two terms are not counteracting one another. This is not true for the large detuning situation as we show next.

The large detuning situation is described by the Hamiltonian of eq.~(\ref{detham}), and we note two important points; the model includes both nearest and next nearest neighbour coupling terms, and the sign of these terms is adjustable by the detuning $\Delta$. The competition between neighbouring terms may give rise to novel phases like {\it charge density waves}~\cite{emil}, {\it supersolids}~\cite{ss} and also to frustration~\cite{j1j2,frust}. The paradigm model showing frustration at a square lattice is the $J_1$-$J_2$ model~\cite{j1j2}, see eq.~(\ref{j1j2}) below. Here, $J_1$ gives the nearest neighbour and $J_2$ the next nearest neighbour Heisenberg anti-ferromagnetic couplings. With $J_1=0$, {\it i.e.} only the couplings along the diagonals are present, the model decouples into two square lattices and the ground state consists of two independent anti-ferromagnetic states in the two sub-lattices. With, instead, $J_2=0$ the model is just the regular nearest neighbour Heisenberg model with the ground state an anti-ferromagnetic state in the full lattice. There is no solution that simultaneously fulfills anti-ferromagnetic order on both sub-lattices and the full lattice. Considering the classical counterpart of the $J_1$-$J_2$ model, {\it i.e.} an Ising model with competing nearest and next nearest couplings, then for $J_2>2J_1$ the system orders such that the two sub-lattices show anti-ferromagnetic order ({\it striped phase}), while for $J_2<2J_1$ anti-ferromagnetic order is established in the full lattice ({\it Neel phase}). At the point $J_2=2J_1$ the system cannot decide for a unique ground state and it gets frustrated with a large number of possible choices.

\begin{figure}[h]
\centerline{\includegraphics[width=6cm]{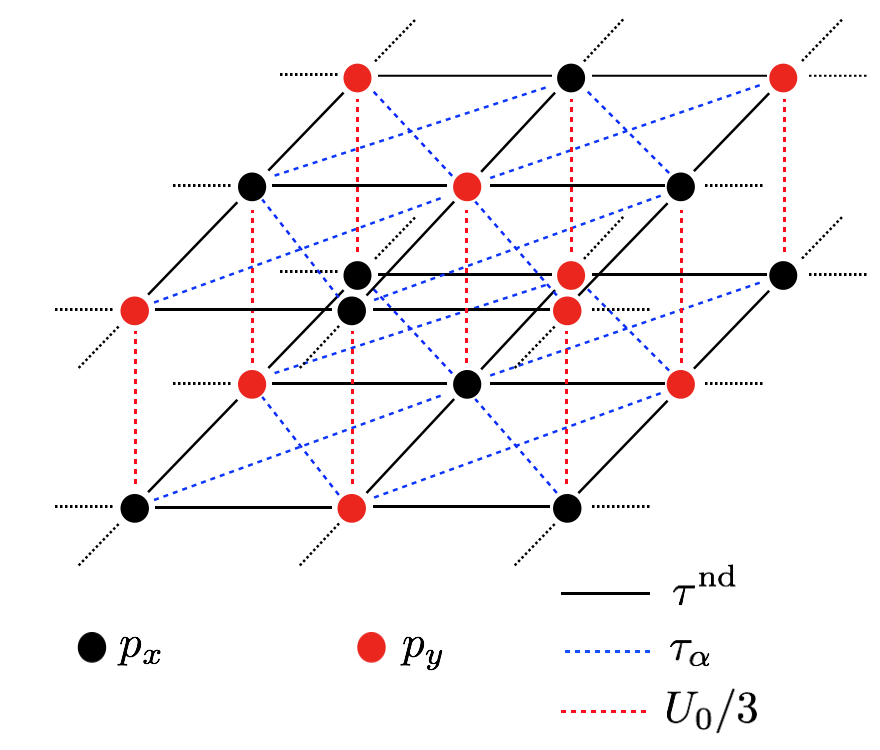}}
\caption{Effective lattice structure of the large detuning model~(\ref{detham}). Instead of double orbital occupancy on every single lattice site, we construct a layered two dimensional lattice. Black dots represent $p_x$-orbitals and red dots $p_y$-orbitals, such that the two orbitals alternate between neighbouring sites. Thus, the solid black lines give the couplings $\tau^\mathrm{nd}$ which swap the orbital state upon tunneling. The dashed blue lines are the diagonal tunnelings $\tau_\alpha$ which keep the orbital state, and dashed red lines are the onsite interaction couplings with strength $U_0/3$. In each two dimensional layer there is a natural sub-lattice structure represented here by either $p_x$- or $p_y$-orbitals, {\it i.e.} black or red dots in each layer mark sub-lattice $A$ and $B$. } \label{fig4}
\end{figure} 

Returning to the present model of eq.~(\ref{detham}). When the onsite interaction is strong we expect the mean-field state to be comprised of vortices and anti-vortices at the different sites. The tunneling terms phase-lock the vortices in the configuration such that it minimizes the total energy. When mimicking the energy functional with respect to the mean-field parameters, provided that $\delta=0$, one finds that the densities of the two orbitals are equal, $n_{x\mathbf{j}}=n_{y\mathbf{j}}\equiv n$. This is expected, but we may note that in higher dimensions this symmetry may indeed be spontaneously broken~\cite{jani}. In the polar representation, $\alpha_{x\mathbf{j}}=\sqrt{n_{x\mathbf{j}}}\exp(i\phi_{x\mathbf{j}})$ and $\alpha_{y\mathbf{j}}=\sqrt{n_{y\mathbf{j}}}\exp(i\phi_{y\mathbf{j}})$, the energy depends on the angles and the global $n$ which we set to unity.  For now we assume the isotropic lattice, $\delta=0$, and when omitting constant terms the mean-field energy becomes ($n=1$)
\begin{equation}\label{mfen}
\begin{array}{lll}
E_\mathrm{MF}[\phi_{\alpha\mathbf{j}}] & = & 
\displaystyle{-2\tau\sum_\alpha\sum_{\{\mathbf{ij}\}_\alpha}\cos(\phi_{\alpha\mathbf{i}}-\phi_{\alpha\mathbf{j}})-2\tau\sum_{\alpha\beta,\,\alpha\neq\beta}\sum_{\langle\mathbf{ij}\rangle}\cos(\phi_{\alpha\mathbf{i}}-\phi_{\beta\mathbf{j}})}\\ \\
& & \displaystyle{+\frac{U_0}{3}\sum_\mathbf{j}\cos(2(\phi_{x\mathbf{j}}-\phi_{y\mathbf{j}}))}.
\end{array}
\end{equation}
This is a two-flavour {\it rotor model} or classical $XY$ model~\cite{rotor}, {\it i.e.} on each site sit two classical rotors which couple onsite to one another with $U_0/3$ and between sites with $2\tau$. It is convenient to introduce an effective lattice for our model where every site hosts instead a single orbital (rotor). This is pictured in Fig.~\ref{fig4}, where we get two layers of planes and on every plane the onsite states alternate from $p_x$ and $p_y$ between every second site. The flavour-changing interaction couples the two layers. In principle, the two orbitals constitute a {\it virtual dimension}, however only two sites long. 

Considering now the case of vanishing interaction, $U_0=0$. We then regain two copies of classical $XY$ models with nearest and next nearest neighbour exchange interactions. The study of frustration in classical $XY$ models on square lattices has a long history~\cite{xy1,xy2,xy3}. It may emerge even for nearest neighbour models on a square lattice provided the tunneling coefficients $\tau_\mathbf{ij}$ on a single plaquette obey a gauge rule which can be viewed as if a certain flux penetrates every plaquette. The un-frustrated regular $XY$ model has zero flux, while the greatest frustration is obtained for half a flux quanta and the corresponding model has been termed {\it fully frustrated} $XY$ or the {\it Villain model}~\cite{xy2}. Historically, frustration in the $XY$ model has mainly been analyzed for nearest neighbour exchanges, and not when next nearest neighbours have been included. When the two terms compete, nearest and next nearest exchange interactions, frustration may occur even when there is no synthetic flux through the plaquette~\cite{xy7,xy8,largeN, heis1}, for the same reason as for the $J_1$-$J_2$ model mentioned above~\cite{j1j2,frust}. Hence, the lattice can be decomposed into two $\sqrt{2}\times\sqrt{2}$ sub-lattices $A$ and $B$ (see also Fig.~\ref{fig4}), and anti-ferromagnetic order cannot simultaneously be fulfilled on the two sub-lattices and the full lattice. With $U_0=0$, the model~(\ref{mfen}) is nothing but a classical $J_1$-$J_2$ model with the quantum spins replaced by classical vector spins $s_{\alpha\mathbf{j}}=(\sin\phi_{\alpha\mathbf{j}},\cos\phi_{\alpha\mathbf{j}})$. For ferromagnetic couplings $\tau>0$, when there is no frustration, it is easy to construct the solution analytically (even for non-zero $U_0$) as demonstrated in Fig.~\ref{fig3} (b). In particular, if we fix the onsite vortex solution at one site, say $(\phi_{x\mathbf{j}},\phi_{y\mathbf{j}})=(0,\pi/2)$, this will determine the vortices at every other sites. In particular, at every second site one finds $(\phi_{x\mathbf{j}},\phi_{y\mathbf{j}})=(0,\pi/2)$, and every other second sites $(\phi_{x\mathbf{j}},\phi_{y\mathbf{j}})=(\pi/2,0)$. Since nothing prevents us from reversing the checkerboard, {\it i.e.} flip vortices to anti-vortices and vice versa, this state is a $\mathbb{Z}_2$-symmetry broken phase. In general, the model hosts two symmetries; one continuous represented by a global phase shift $\phi_{\alpha\mathbf{j}}\rightarrow\phi_{\alpha\mathbf{j}}+\nu$ for arbitrary $\nu$ (this is the corresponding symmetry deriving from particle conservation), and a chiral $\mathbb{Z}_2$-symmetry $\phi_{\alpha\mathbf{j}}\rightarrow-\phi_{\alpha\mathbf{j}}$. The order of the ground state of the $XY$ model with nearest and next nearest anti-ferromagnetic ({\it i.e.} frustrated) neighbour exchange interactions is determined by the relative strengths between the two couplings $J_1$ and $J_2$. When the next nearest neighbour interaction dominates, $|J_2|>|J_1|$, the system builds-up independent anti-ferromagnetic order on each sub-lattice $A$ and $B$~\cite{xy7,xy8}. When instead the nearest neighbour dominates, $|J_1|>|J_2|$, the anti-ferromagnetic order is established in the full lattice. At the point $|J_1|=|J_2|$ the system is fully frustrated.

Our model is, of course, more interesting for anti-ferromagnetic couplings $\tau<0$ when it is fully frustrated~\cite{com2}. 
The fully frustrated $XY$ model is especially appealing due to the spin glass phase existing despite lack of any disorder that manifestly breaks translational symmetry~\cite{xy1,xy3}. For the Villain model, the nature of the transition from a disordered to a glass phase has been throughly discussed. The continuous symmetry cannot be spontaneously broken according to the {\it Mermin-Wagner theorem}~\cite{auerbach}, and the corresponding transition should instead be of the {\it Kosterlitz-Thouless} type. However, the discrete $\mathbb{Z}_2$-symmetry can indeed be spontaneously broken and could give rise to an Ising type transition~\cite{xy4}. Three scenarios emerge; whether ($i$) it is a Kosterlitz-Thouless transition followed by an Ising one, ($ii$) a mixture of the two occurring simultaneously for the same critical temperature, or ($iii$) a transition belonging to a new universality class, see Refs.~\cite{xy3}. The consensus seems to be that there are actually two nearby transitions, {\it i.e.} alternative ($i$). It has been argued that the state of the system in the narrow window between the two transitions is a chiral spin liquid~\cite{xy5,obd}.  

At zero temperature a glass phase appears, and due to the chiral symmetry the ground state is doubly degenerate. Even though our model of eq.~(\ref{mfen}) is not exactly a classical $XY$ model due to the layered structure, the two models should show the same generic features. The relevant question is what fluctuations, either thermal or quantum, do to the phases. Small perturbations to the degenerate ground state could lift the degeneracy and cause long range order, the so called {\it order-by-disorder} mechanism~\cite{orderdisorder}. For the $XY$ model with nearest and next nearest neighbour exchange interactions, in the phase dominated by next nearest neighbour interaction it was shown using spin wave theory that these fluctuations order the state such that the anti-ferromagnets in the two sub-lattices become collinear, {\it i.e.} the relative phase between two neighbouring sites is either $0$ or $\pi$~\cite{xy7}. We are, however, interested in the point when $J_1=J_2$ and there is no preferred order between the sub-lattices and the full one. One possibility, see also next Subsection, is a spin liquid phase that survives in the vicinity of the point $J_1=J_2$~\cite{heis1}. A large-$N$ expansion indicated that such a spin liquid phase should only exist at exactly the degeneracy point, and not in its vicinity~\cite{largeN}.

\begin{figure}[h]
\centerline{\includegraphics[width=9cm]{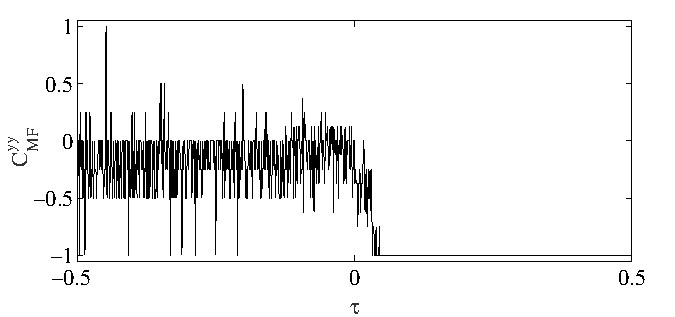}}
\caption{The mean-field correlator $\mathcal{C}_\mathrm{MF}^{yy}$ of Eq.~(\ref{corr}) as a function of the coupling $\tau$, ranging from negative to positive values. The lattice is considered isotropic and for the numerics we have used a $4\times4$ lattice with periodic boundary conditions, and furthermore we have set $U_0=1$. For positive tunneling amplitudes $\tau$, the system orders in an anti-ferromagnetic phase marked by $\mathcal{C}_\mathrm{MF}^{yy}=-1$. However, when we cross over to negative $\tau$ the correlator starts to fluctuate greatly from one $\tau$-value to the next. Such fluctuations signal that there are several mean-field solutions with roughly the same energy and the numerical algorithm randomly picks one of them. If the lattice were large enough we would expect that $\mathcal{C}_\mathrm{MF}^{yy}=0$ for every numerical simulation. } \label{fig5}
\end{figure}

We next verify the presence of frustration numerically by finding the ground state for a $4\times4$ lattice. The anti-ferromagnetic order obtained when $\tau>0$ is reflected in the nearest neighbour ($|\mathbf{i}-\mathbf{j}|=1$) correlator
\begin{equation}\label{corry}
C^{yy}(\mathbf{i},\mathbf{j})=\langle\hat S_\mathbf{i}^Y\hat S_\mathbf{j}^Y\rangle,
\end{equation}
which is negative for the anti-ferromagnetic phase (note that the Schwinger boson operator $\hat S_\mathbf{j}^Y$ is identical to the angular momentum operator $\hat L_{z\mathbf{j}}$ defined in Eq.~(\ref{angop})). In the mean-field approximation we have (with $n=1$) 
\begin{equation}
C_\mathrm{MF}^{yy}(\mathbf{i},\mathbf{j})=\sin\left(\phi_{x\mathbf{j}}-\phi_{y\mathbf{j}}\right)\sin\left(\phi_{x\mathbf{i}}-\phi_{y\mathbf{i}}\right),
\end{equation}   
which for the anti-ferromagnetic phase becomes $C_\mathrm{MF}^{yy}(\mathbf{i},\mathbf{j})=-1$. We can go on and define the average correlator
\begin{equation}\label{corr}
\mathcal{C}_\mathrm{MF}^{yy}=\frac{1}{2N}\sum_{\langle\mathbf{i}\mathbf{j}\rangle}C_\mathrm{MF}^{yy}(\mathbf{i},\mathbf{j}),
\end{equation}
where $N$ is the total number of sites, and the sum is over all nearest neighbours in the full lattice. It is clear that $-1\leq\mathcal{C}_\mathrm{MF}^{yy}\leq+1$ with $\mathcal{C}_\mathrm{MF}^{yy}=-1$ for the anti-ferromagnetic phase and $\mathcal{C}_\mathrm{MF}^{yy}=+1$ for the ferromagnetic phase.

In Fig.~\ref{fig5}, we display the numerical results for the correlator  $\mathcal{C}_\mathrm{MF}^{yy}$ showing how it depends on the coupling $\tau$. At positive couplings we verify the analytically predicted anti-ferromagnetic order which persists down to $\tau=0$ (apart from numerical fluctuations close to $\tau=0$). For negative $\tau$ the numerics shows large fluctuations from one simulation to the next meaning that the numerical minimization algorithm finds very different ground states. This is indeed a smoking gun of classical frustration. We have also verified that the same behaviour is retainable when the densities of $p_x$- and $p_y$-orbitals are allowed to vary. In particular, as stated above we do not find any breaking of the balanced population between the two orbitals. We expect that the fluctuations in the frustrated phase decreases with the system size, {\it i.e.} in the thermodynamic limit $\mathcal{C}_\mathrm{MF}^{yy}=0$ for $\tau<0$. 

In the experiments~\cite{hemmerich1,hemmerich2}, with $\Delta=0$, the order of the condensate could be determined in a time-of-flight measurement. More precisely, the detection verified a complex order parameter and thereby a broken time-reversal symmetry. In the present model with a large detuning $\Delta$, the same type of measurement will give direct fingerprints of the phases. The ferromagnetic case with $\tau>0$ should produce similar time-of-flight images like in~\cite{hemmerich1,hemmerich2}, while in the frustrated case the images will look very different due to the more irregular global phase order of the condensate. Another experimentally relevant question is how to realize the state at the first place. For anti-ferromagnetic coupling terms, which is crucial for achieving frustration, the $s$-orbital sites are lower in energy than the $p$-orbital ones. In other words, the atoms reside on the higher excited sites and could in principle relax down to lower lying energy sites. However, this relaxation can become very long for the same reason as why the life-time of the original Hemmerich $s$-$p$ hybradised system is surprisingly long~\cite{hemmerich1}. Namely, relaxation processes of atoms from the $p$-bands into other bands are not energetically allowed, and these states are thereby metastable.

\begin{figure}[h]
\centerline{\includegraphics[width=9cm]{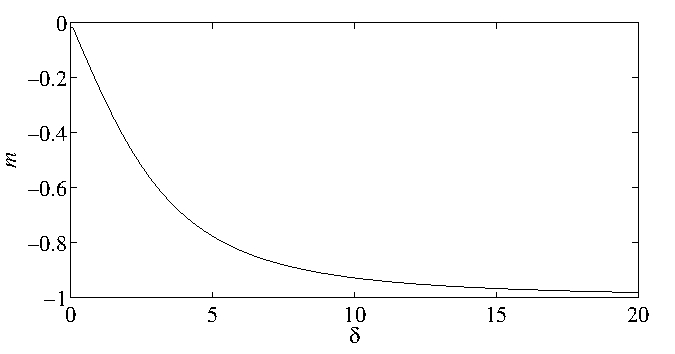}}
\caption{The magnetization $m$ as a function of $\delta$ for $U_0=1$ and $\tau=1$. The transition from an anti-ferromagnetic phase in the $y$-direction to a polarized phase in the $z$-direction is smooth, {\it i.e.} there is no true critical point separating the two extremes. As in the previous figure, the lattice is $4\times4$ and the ground state is found numerically.  } \label{fig6}
\end{figure}

We end this section by mentioning the behaviour for non-zero detuning $\delta$, {\it i.e.} when one of the two orbitals is energetically favoured. Thus, we cannot assume a balanced population between the two orbitals, but must minimize the energy functional also with respect to $n_{x\mathbf{j}}$ and $n_{y\mathbf{j}}$. In the anti-ferromgnatic regime ($\tau>0$) we find a smooth transition to a polarized phase, as depicted in Fig.~\ref{fig6}. The figure shows the $\delta$ dependence of the scaled magnetization $m=M/N$, where $N$ is the number of sites and 
\begin{equation}
M=\sum_\mathbf{j}\left(n_{x\mathbf{j}}-n_{y\mathbf{j}}\right).
\end{equation}
The behaviour is conceptually different for negative $\tau$ where, for $\delta=0$, the classical state is frustrated. Here, a small non-zero $\delta$ destabilizes the ground state and it gets fully polarized. This is easily understood from the fact that there is a great set of (quasi) degenerate ground states and a small perturbation may easily lift this degeneracy. The size of $\delta$ for this to occur is determined by the interaction amplitude $U_0$ which favours the onsite vortex solutions. In the thermodynamic limit we expect that any non-zero $\delta$ will cause the system to polarize, and the frustration survives only at the symmetry point $\delta=0$. The system thereby shows a first order PT at $\delta=0$.

\subsection{Mott insulating phase diagram}
As described in Subsec.~\ref{sec2.3}, the insulating regime is analyzed by a $t/U_0$-expansion that produces an effective Hamiltonian on the form of eq.(\ref{HamMott}). This is part of a larger group of Hamiltonians that can be expressed as
\begin{equation}
  \hat H = \sum_{\textbf{j}}\sum_{\alpha}h^{\alpha}\hat S_\mathbf{j}^\alpha+\sum_{[\textbf{i}\textbf{j}]_{\sigma}}\sum_{\alpha}J^{\alpha}_{\sigma}\hat S^{\alpha}_{\textbf{i}}\hat S_{\textbf{j}}^{\alpha}
\end{equation}
where the second sum includes the bonds of interest (nearest neighbor, next nearest neighbor, etc). The most famous model of this kind is probably the $J_1$-$J_2$ model 
\begin{equation}\label{j1j2}
\hat H_{J_1J_2}=J_1\sum_{\langle\mathbf{ij}\rangle}\hat S_\mathbf{i}\cdot\hat S_\mathbf{j}+J_2\sum_{\{\mathbf{ij}\}}\hat S_\mathbf{i}\cdot\hat S_\mathbf{j},
\end{equation}
where $\hat S_\mathbf{i}=(\hat S_\mathbf{i}^X,\hat S_\mathbf{i}^Y,\hat S_\mathbf{i}^Z)$. As mentioned above, for couplings supporting anti-ferromagnetic order a competition arises which results in classical frustration at the point $J_1=2J_2$~\cite{j1j2}. For the classical system\footnote{In the classical system we only consider Ising spins $s=\pm1$, or in other words only $\hat S_\mathbf{i}^Z$ operators and not $\hat S_\mathbf{i}^X$ nor $\hat S_\mathbf{i}^Y$.}, this point separates the Neel and striped phases. There is an ongoing debate whether quantum fluctuations (including couplings between $\hat S_\mathbf{i}^X$ and/or $\hat S_\mathbf{i}^Y$) can open up for a QSL phase in the vicinity of the degeneracy point $J_1=2J_2$~\cite{sql1,sql2,sql3,sql4}. The consensus is that indeed an intermediate phase emerges, and the estimated range for this quantum phase is $0.42\lesssim J_2/J_1\lesssim0.62$, as quantum fluctuations are taken into account. State-of-the-art numerical simulations indicate that this is a QSL, either a gapped~\cite{sql1,sql2}, or a gapless $\mathbb{Z}_2$ QSL~\cite{sql3}. However, a more recent work, using renormalization group arguments, suggests instead that this phase is not a true QSL, but rather a so called {\it plaquette valence-bond} phase~\cite{sql4}.

\begin{figure}[h]
\centerline{\includegraphics[width=15cm]{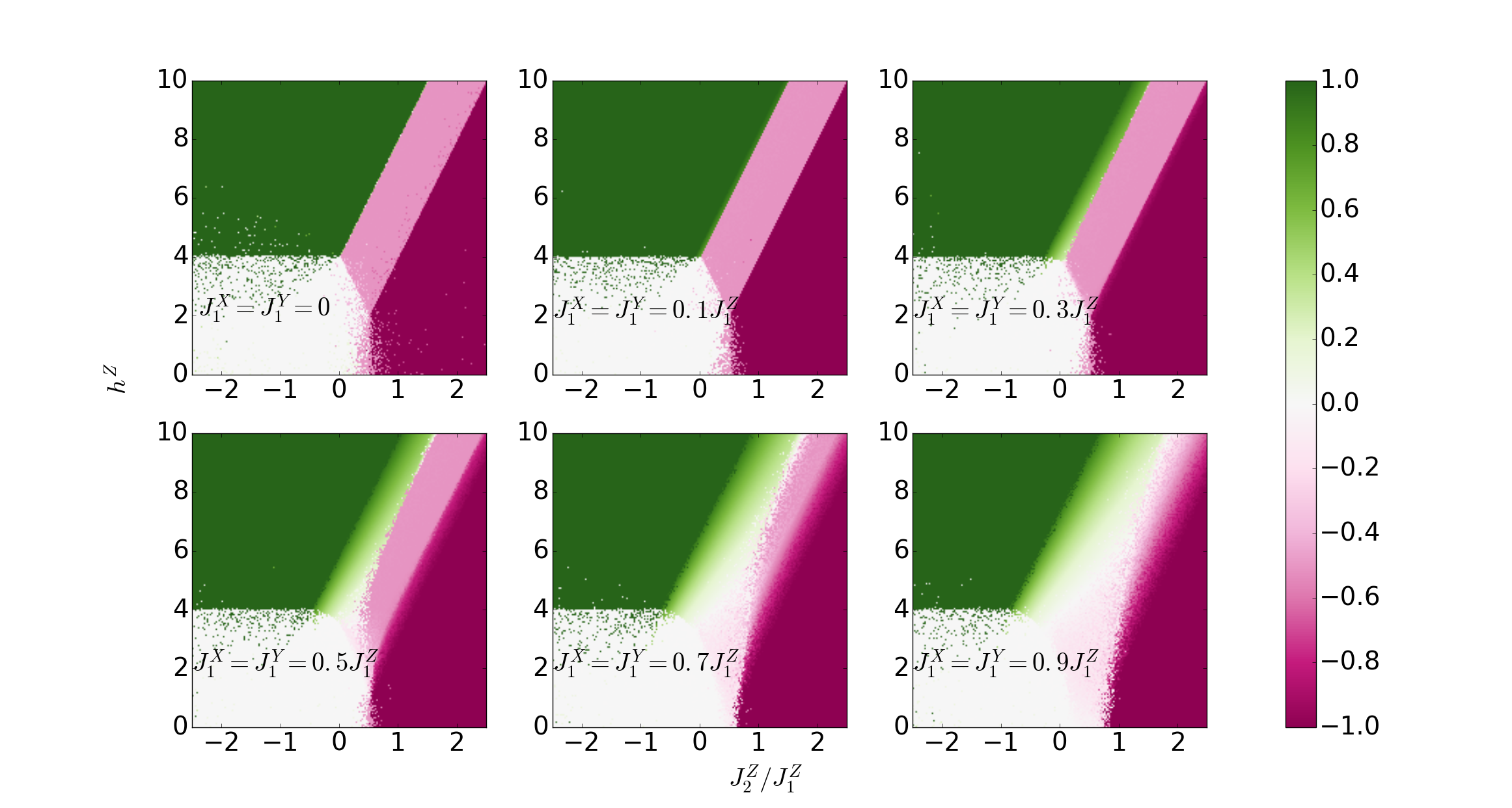}}
\caption{The mean-field phase diagram of the model of eq.~(\ref{HamMott}), here characterized by the full correlator $\mathcal{C}_\mathrm{full}$ defined in~(\ref{corr}) and with $J_1^{X}=J_1^{Y}$. Each plot represents a different coupling strength of $J_1^{X}=J_1^{Y}$ as indicated by the textin the figures. In the case of $J_1^{X}=J_1^{Y}=0$ we see four distinct phases, the ferromagnetic phase (green), the anti-ferromagnetic Neel phase (white), the striped anti-ferromagnetic phase (dark pink), and the disordered phase (pink). In this (classical) limit all phase transitions are first first order. As we consider non-zero couplings $J_1^{X}=J_1^{Y}$ a fifth phase appears (light green/pink), this phase grows with increasing couplings $J_1^{X}=J_1^{Y}$, and all PTs except for the one between the anti-ferromagnetic Neel phase and the ferromagnet phase, and the anti-ferromagnetic Neel and the new phase, appears to be second order. The new phase survives also for zero field, $J^Z=0$, provided that $J_1^{X}=J_1^{Y}$ is large enough. The dispersed dots are numerical errors for which the simulations are not capable of finding the true ground state.} \label{PD_iso}
\end{figure}

Comparing eq.~(\ref{HamMott}) to the full $J_1$-$J_2$ model it is clear that our model does not comprise all the six terms of the $J_1$-$J_2$ model since the next nearest neighbor couplings of the $X$ and $Y$ terms are absent. In that sense our model is less symmetric. Moreover, the coupling strengths of the three nearest neighbour terms are different from one another. In addition, any anisotropy in the model will induce an effective field in the $Z$-direction, which we thereby include in our model. We will actually see that the field is of great importance. It appears already at zeroth order in the perturbation and the field strength $h^Z$ can to a good approximation be considered as a free parameter~\cite{fernanda}. Surprisingly, to the best of our knowledge, we do not know of any work that explores the quantum $J_1$-$J_2$ model in the presence of a field. However, the classical $J_1$-$J_2$ model with a field has been analyzed in Ref.~\cite{classicalJ1J2}. The phase diagram is known to consist of four phases; the aforementioned anti-ferromagnetic Neel ordered phase and anti-ferromagnetic striped ordered phase that survive at zero field $h=0$, and a ferromagnetic (or polarized) ordered phase that appears for sufficiently strong fields $h^Z$, and finally a {\it disordered phase} that also only exists for non-vanishing fields. The locations of the phase boundaries can be found analytically, and naturally when considering classical Ising spins at zero temperature all transitions are first order. The phase diagram for the classical model is depicted in fig.~\ref{PD_iso} (a).  

To distinguish between these four phases we define
\begin{equation}\label{corr}
\mathcal{C}_\mathrm{full} = \frac{1}{N}\left(\sum_{\textbf{i}}\langle S^{Z}_\textbf{i}\rangle + \frac{1}{2}\sum_{\langle\textbf{i},\textbf{j}\rangle}C^{zz}(\textbf{i},\textbf{j}) + \frac{1}{2}\sum_{\left\{\textbf{i},\textbf{j}\right\}}C^{zz}(\textbf{i},\textbf{j})\right),
\end{equation}
what we call the \textit{full correlator} and where, in analogy to eq.~(\ref{corry}),
\begin{equation}
C^{zz}(\textbf{i},\textbf{j}) = \langle S_{\textbf{i}}^Z S_{\textbf{j}}^Z\rangle.
\end{equation}
The full correlator is restricted by $-1\leq\mathcal{C}_\mathrm{full}\leq1$, and we note that the ferromagnetic phase is characterized by $\mathcal{C}_\mathrm{full}=1$, the anti-ferromagnetic Neel phase by $\mathcal{C}_\mathrm{full}=0$, the anti-ferromagnetic striped phase by $\mathcal{C}_\mathrm{full}=-1$, and finally the disordered phase by $\mathcal{C}_\mathrm{full}=-1/2$. In order to investigate whether quantum fluctuations may affect the phase diagram in fig.~\ref{PD_iso} (a) of the classical model, we apply the simplest version of mean-field theory to our model. Various mean-field approaches have been employed when studying the $J_1$-$J_2$ model~\cite{j1j2mf}. In general, we cannot expect a quantitatively, and sometimes not even a qualitatively, correct description from mean-field results, but they can signal certain properties as we will see. Adopting the approach of the previous subsection, the full system state is factorized between the neighbours, and single site spins are parametrized as $\left(\hat S_\mathbf{i}^X,\hat S_\mathbf{i}^Y,\hat S_\mathbf{i}^Z\right)=\left(S\sin\theta_\mathbf{i}\cos\phi_\mathbf{i},S\sin\theta_\mathbf{i}\sin\phi_\mathbf{i},S\cos\theta_\mathbf{i} \right)$, where $S=1/2$ is the spin. This is analogous to assigning spin coherent states to every site, just like we did above but with boson coherent states. The energy $E\left[\theta_\mathbf{i},\phi_\mathbf{i}\right]$ is then numerically minimized with respect to the polar and azimuthal angles $\theta_\mathbf{i}$ and $\phi_\mathbf{i}$ respectively. The results of such a treatment are presented in fig. \ref{PD_iso}. In frame (a) we reproduce the classical result~\cite{classicalJ1J2} up to  numerical errors (seen as `scatters' in the vicinities of the phase boundaries). In particular, the four phases mentioned above are clearly visible. As $J_1^{X}$ ($=J_1^{Y}$) becomes non-zero, the model is no longer integrable and quantum fluctuations also set in. First we see that the transitions between the disordered phase and the ferromagnetic and striped phases turn continuous. What is more interesting is that the mean-field results suggest that a new phase emerges in the disordered phase (light green/pink). When $J_1^{X}$ ($=J_1^{Y}$) is increased further this phase extends down to the zero field limit. In particular, this happens around the classical highly degenerate point $J_1^{Z}=2J_2^{Z}$. It seems that the boundaries for this phase approximately agree with the aforementioned limits $0.42\lesssim J_2^{Z}/J_1^{Z}\lesssim0.62$ for the possible QSL phase. Thus, by assuming that this phase is the same as the phase at the axis $h^Z=0$ our results indicate that the intermediate phase survives the presence of a field. 

\begin{figure}[h]
\centerline{\includegraphics[width=11cm]{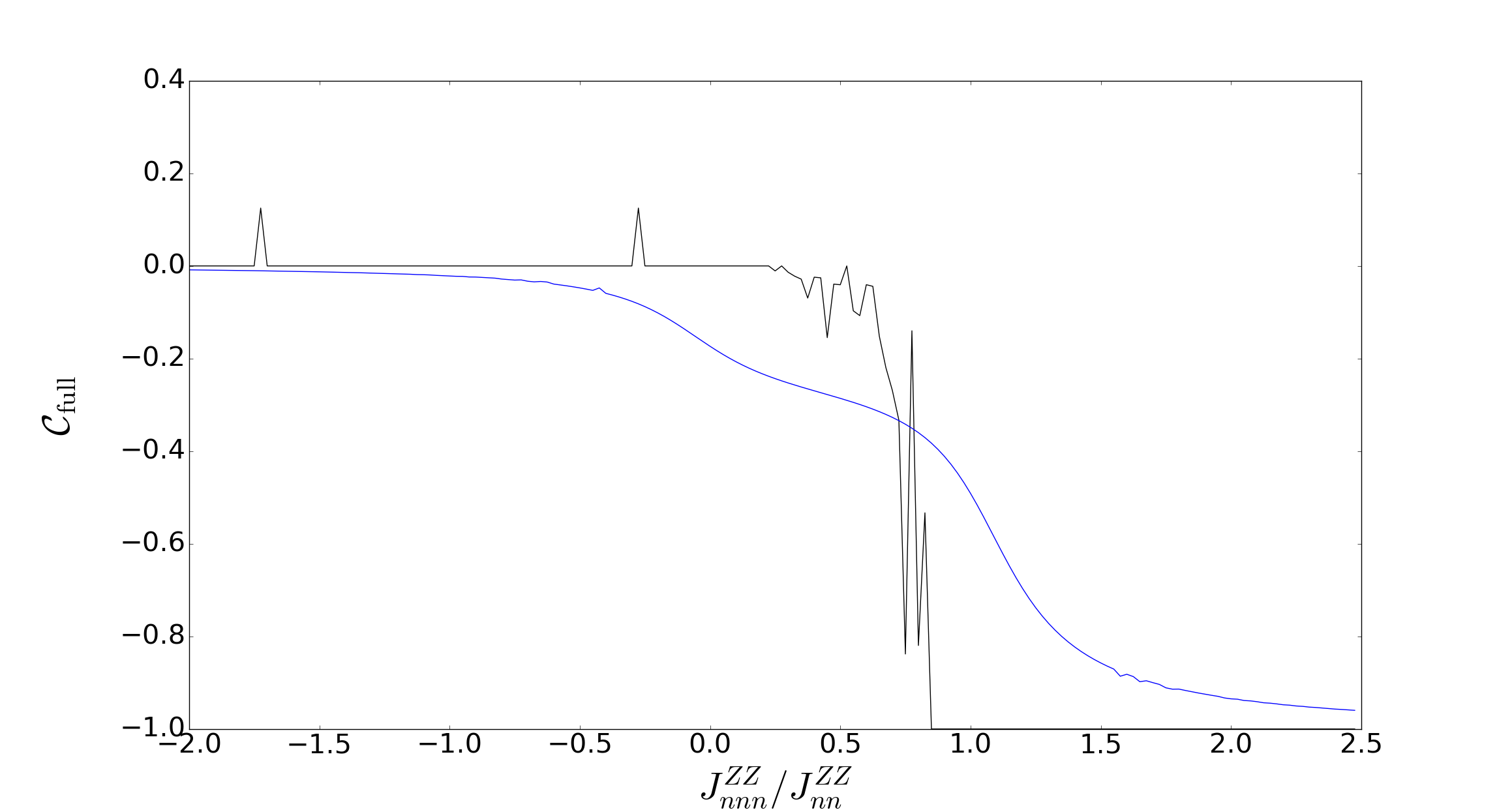}}
\caption{ The full correlator of eq.~(\ref{corr}) for zero field $h^Z$. The result obtained by mean-field is shown as the black line, while the result from exact diagonalisation of a $4\times4$ lattice is represented by the blue line. Remember that for the anti-ferromagnetic Neel phase $\mathcal{C}_\mathrm{full}=0$, while in the striped anti-ferromagnetic phase $\mathcal{C}_\mathrm{full}=-1$. The range in between these two phases marks the intermediate new phase. For such a small lattice as this, the finite size effects dominate and because of this the exact diagonalisation over estimates the intermediate phase. The results of the exact diagonalisation indicates that in the intermediate phase a plateau is formed, which could hint the presence of a new intermediate phase. The parameters used for this figure are $J_1^{X}= J_1^{Y}=0.9$. The sudden jumps of the black line are again numerical artifacts when the code finds a local minimum and not the global minimum of the energy functional. } \label{MFED_iso}
\end{figure}

We should point out that the mean-field study cannot be expected to give definite answers. For example, it does not support nor rule out the possibility that the new phase is a QSL. However, it is an interesting observation that the boundaries for the new phase coincide with those suggested for a QSL at zero field. Unfortunately, exact diagonalisations are limited to very small lattices which are not capable of extracting long-range correlations. Nevertheless, in fig.~\ref{MFED_iso} we compare the mean-field results for $\mathcal{C}_\mathrm{full}$ with those of an exact diagonalisation for a $4\times4$ lattice. Expectedly, the finite-size effects cause the crossover region between the two anti-ferromagnetic phases to be more extended. However, it is interesting to see tendencies of a plateau forming in the regime where the mean-field results predict a new phase. We take this as another indicator of the existence of a fifth phase. 

One further observation of fig.~\ref{PD_iso} is the extension of the new phase into the regime of negative ratios $J_2^{Z}/J_1^{Z}$. For a field $h^Z=4$, the phase seems to survive down to $J_2^{Z}/J_1^{Z}\approx-1$ when $J_1^{X}$ ($J_1^{Y}$ ) is of the same order as $J_1^{Z}$. Returning to the actual effective model of eq.~(\ref{HamMott}) that describes the Mott$_1$ insulating phase of our bipartite optical lattice model we note that identically $J_2^{Z}/J_1^{Z}=-1/2$, and thus we sit on the corresponding vertical line in the phase diagram upon varying $h^Z$. Naturally, the most interesting phase is the new intermediate phase, which may actually be a QSL or at least host exotic quantum correlations, and to reach this phase experimentally one would need to tune the field such that $h^Z\approx4$.

\section{Conclusion}
We have discussed phases of a two-dimensional bipartite optical lattices that may emerge in various parameter regimes. The feature leading to these novel phases is the alternation between $s$- and $p$-orbital sites. The sites hosting $s$-orbital atoms induce an effective coupling between $p$-orbital atoms, resulting in competing nearest and next nearest neighbour interactions. For effective anti-ferromagnetic coupling terms the system becomes frustrated.

For the Mott insulating phase with a single atom per site, we used perturbation theory to derive an effective spin-1/2 model for the $p$-orbital atoms. The resulting Hamiltonian is similar to the well known $J_1$-$J_2$ model that has been thoroughly studied in the past. As a candidate model for the realization of QSL phases, we discussed the possibilities to find also related phases in our system. In particular, our mean-field analysis suggested the appearance of a new phase whose phase boundaries seemed to agree with those of the predicted QSL. This new phase survived a non-vanishing field (which would automatically arise in an anisotropic lattice), and we argued that in order to explore this phase experimentally one should consider a non-zero field. 

In the superfluid phase we again derived an effective model for the $p$-orbital atoms, but this time by assuming a large energy detuning between $s$- and $p$-orbital atomic states. In this regime we can adiabatically eliminate the $s$-orbital sites/states. While in the insulating regime a mean-field approach could be questioned, in the superfluid phase we believe that it should give a much better description of the physical system. Within this framework we derive a semi-classical model showing great resemblance with a classical $XY$ model. Our system becomes fully frustrated in the anti-ferromagnetic regime, and we discuss whether this could give rise to glass or liquid phases.

To numerically distinguish a true QSL phase from other possible phases is very hard~\cite{spinliq}. These are points that would need further investigation, {\it i.e.} do we have true QSL phases, and if not what are these phases? And how would they be experimentally probed? As mentioned above, frustration in the superfluid phase should manifest in time-of-flight measurements. To measure quantum correlations one would need more sophisticated methods like single-site detection~\cite{sings}.

\ack We thank Axel Gagge, Andreas Hemmerich and Themistoklis Mavrogordatos for insightful discussions. The Knut and Alice Wallenberg foundation (KAW) and the Swedish research council (VR) are acknowledged for financial support. 

\appendix
\section{Adiabatic elimination of $s$-orbital degrees-of-freedom}\label{appa}
Under the assumption that $|\Delta|$ is mach larger than the remaining parameters, the $s$-orbital atoms will be slaved to the $p$-orbital ones, {\it i.e.} the $s$-orbital atoms evolve with the smallest characteristic time-scale. Under this circumstance they adiabatically follow the evolution of the other atoms, and it is thereby legitimate to adiabatically eliminate the $s$-orbital degrees-of-freedom. Following the standard procedure~\cite{gardiner} we start from the Heisenberg equations-of-motions as obtained from the full Hamiltonian, Eqs.~(\ref{ham1})-(\ref{ham3}), to derive
\begin{equation}
\begin{array}{lll}
\partial_t\hat a_{s\mathbf{i}} & = & -i\Delta\hat a_{s\mathbf{i}}-iU_{ss}\hat n_{s\mathbf{i}}\hat a_{s\mathbf{i}}\\ \\
& & it_x\left(\hat a_{x(\mathbf{i+1}_x)}+\hat a_{x(\mathbf{i-1}_x)}\right)+t_y\left(\hat a_{y(\mathbf{i+1}_y)}+\hat a_{y(\mathbf{i-1}_y)}\right),\\ \\
\partial_t\hat a_{x\mathbf{j}} & = &it_x\!\left(\hat a_{s(\mathbf{j+1}_x)}\!+\hat a_{s(\mathbf{j-1}_x)}\right)+\mathrm{interaction\,\,terms},\\ \\
\partial_t\hat a_{y\mathbf{j}} & = & it_y\!\left(\hat a_{s(\mathbf{j+1}_y)}\!+\hat a_{s(\mathbf{j-1}_y)}\right)+\mathrm{interaction\,\,terms}.
\end{array}
\end{equation}
Here we have introduced the notation $\mathbf{i\pm1}_x$ for the horizontal neighbouring sites to site $\mathbf{i}$, and  $\mathbf{i\pm1}_y$ the same but in the vertical direction. The steady-state solution for the $s$-orbitals is obtained from $\partial_t\hat a_{s\mathbf{i}}=0$, and we make the further assumption that the $\mathcal{S}$-sites are initially empty, such that for all times $\langle\hat n_{s\mathbf{i}}\rangle\ll1$ and we therefore neglect the shift deriving from onsite interaction on these sites. The steady state solution then becomes
\begin{equation}\label{ss}
\hat a_{s\mathbf{i}}^{\mathrm{(ss)}} = \frac{t_x}{\Delta}\left(\hat a_{x(\mathbf{i+1}_x)}+\hat a_{x(\mathbf{i-1}_x)}\right)+\frac{t_y}{\Delta}\left(\hat a_{y(\mathbf{i+1}_y)}+\hat a_{y(\mathbf{i-1}_y)}\right).
\end{equation}
When substituting this expression for the $s$-orbital operators in the equations-of-motion for $\hat a_{x\mathbf{j}}$ we obtain a series of terms. Apart from the unaffected interaction terms, all of these represent different two-step processes involving $x$ and $y$ orbital atoms, and hence only operators on the $\mathcal{P}$-sites. We write
\begin{equation}
\partial_t\hat a_{x\mathbf{j}}=\hat f_2+\hat f_1
\end{equation}
where we have divided the terms into different categories. The first
\begin{equation}\label{a4}
\hat f_2=i\tau_x\left(\hat a_{x(\mathbf{j+2}_x)}+\hat a_{x(\mathbf{j-2}_x)}+2\hat a_{x\mathbf{j}}\right)
\end{equation}
gives horisontal (diagonal in the $\mathcal{P}$-sub-lattice) tunneling of $p_x$-orbital atoms with (as defined in the main text) the amplitude $\tau_x=|t_x|^2/\Delta$. In the resulting lattice for the $p$-orbital atoms, these terms describe next nearest neighbour tunnelings. In addition, the last term in the above expression~(\ref{a4}) represents an onsite energy shift.  Similarly, there are terms describing nearest neighbour tunneling in the $\mathcal{P}$-sub-lattice, {\it i.e.}
\begin{equation}
\hat f_1=i\tau_{xy}\left(\hat a_{y\mathbf{j+1}_x+\mathbf{1}_y}+\hat a_{y\mathbf{j+1}_x-\mathbf{1}_y}+\hat a_{y\mathbf{j-1}_x+\mathbf{1}_y}+\hat a_{y\mathbf{j-1}_x-\mathbf{1}_y}\right),
\end{equation}
with the amplitude $\tau_{xy}=t_xt_y/\Delta$. Equivalently, one finds the corresponding equations for $\partial_t\hat a_{y\mathbf{j}}$. From these equations it is straightforward to derive an effective Hamiltonian for the $p$-orbital atoms via
\begin{equation}
\partial\hat a_{\alpha\mathbf{j}}=-i\left[\hat a_{\alpha\mathbf{j}},\hat H_\mathrm{eff}\right].
\end{equation}

\section{Derivation of the effective Hamiltonian in the Mott phase}\label{appb}
The Hamiltonian in the Mott phase is obtained by considering the tunneling part as a perturbation to the full Hamiltonian. Here the fixed number of atoms per lattice site is effectively handled by dividing the Hilbert space of the eigenvalue problem into two orthogonal subspaces using the projection operators $\hat{P}^{2}=\hat{P}$ and $\hat{Q}^{2}=\hat{Q}$ with $\hat{P}+\hat{Q}=1$. $\hat{P}$ projects onto the subspace $\mathcal{H}_{P}$ where all lattice sites are occupied with one atom, and $\hat{Q}$ projects onto the complementary subspace $\mathcal{H}_{Q}$. The eigenvalue problem may be written as
\begin{equation}
 \hat{H} \left(\hat{Q}+\hat{P}\right)\psi=\left(\hat{H}_{K}+\hat{H}_{U}\right)\left(\hat{Q}+\hat{P}\right)\psi= E\psi,
\end{equation}
where $\hat H_{K}$ is the kinetic part of the Hamiltonian, and $\hat H_{U}$ is the interaction part of the Hamiltonian. This leads to an effective Hamiltonian $\hat H_1$ in the $\mathrm{Mott}_{1}$ phase ({\it i.e.} the insulating phase with unit density)~\cite{fernanda}
\begin{equation}
\hat{H}_{1}\psi=-\hat{P}\hat{H}_{K}\hat{Q}\frac{1}{\hat{Q}\hat{H}\hat{Q}-E}\hat{Q}\hat{H}_{K}\hat{P}\psi.
\end{equation} 
This expression is exact and serves as the starting point for treating the tunneling perturbatively, {\it i.e.} one expands $\frac{1}{\left(\hat{Q}\hat{H}\hat{Q}-E\right)}$. Making the approximation $\frac{1}{\hat{Q}\hat{H}\hat{Q}-E}\approx \frac{1}{\hat{H}-E}$ it follows that
\begin{equation}
\frac{1}{\hat{Q}\hat{H}\hat{Q}-E}\approx\frac{1}{\hat{H}_{U}}\frac{1}{\left( 1+\hat{H}_{U}^{-1}\left(\hat{H}_{K}-E\right)\right)}.
\end{equation}
As the tunneling coefficient is much smaller than the interaction coefficient one may expand around $\hat{H}_{U}^{-1}\left(\hat{H}_{K}-E\right)$.To include terms up to fourth order in the tunneling parameter correspons to expanding $\hat{K}$ to second order
\begin{equation}
 \hat{K}\equiv\frac{1}{\hat{H}_{U}}\left[1+\frac{1}{\hat{H}_{U}}\left(\hat{H}_{K}-E\right)+\frac{1}{\hat{H}_{U}}\left(\hat{H}_{K}-E\right)\frac{1}{\hat{H}_{U}}\left(\hat{H}_{K}-E\right)\right],
\end{equation}
\subsection{Bipartite lattice structures in the Mott phase}
In any Mott phase it is clear that only even terms in the perturbative expansion will be non-zero. For a bipartite lattice one needs to include fourth-order tunneling processes in order to couple two lattice sites that both support the same type of orbital states. For a better understanding of these processes, we consider two generic transitions, one starting in an $\mathcal{S}$-site, which we label $\mathcal{C}^{\mathcal{S}}$, and one starting in a $\mathcal{P}$-site denoted $\mathcal{C}^{\mathcal{P}}$; 
\begin{equation}
\begin{array}{lll}
 \hat\mathcal{C}^{\mathcal{P}}&=&t_{\alpha }t_{\beta}t_{\alpha'}t_{\beta'} \hat{a}_{\beta_{l}'}^{\dagger}\left(\hat{a}_{s_{k}}\hat{K}^{\mathcal{S}}_{k}\hat{a}_{s_{k}}^{\dagger}\right)\left(\hat{a}_{\alpha_{j}}\hat{K}^{\mathcal{P}}_{l}\hat{a}_{\beta_{j}}^{\dagger}\right)\left(\hat{a}_{s_{i}}\hat{K}^{\mathcal{S}}_{i}\hat{a}_{s_{i}}^{\dagger}\right)\hat{a}_{\alpha_{l}'},
 \label{contribution1}\\ \\
  \hat\mathcal{C}^{\mathcal{S}}&=&t_{\alpha }t_{\beta}t_{\alpha'}t_{\beta'} \hat{a}_{s_{i}}^{\dagger}\left(\hat{a}_{\alpha_{l}'}\hat{K}^{\mathcal{P}}_{l}\hat{a}_{\beta_l'}^{\dagger}\right)\left(\hat{a}_{s_{k}}\hat{K}^{\mathcal{S}}_{k}\hat{a}_{s_{k}}^{\dagger}\right)\left(\hat{a}_{\alpha_{j}}\hat{K}^{\mathcal{P}}_{j}\hat{a}_{\beta_j}^{\dagger}\right)\hat{a}_{s_{i}}.\label{contribution2}
\end{array}
\end{equation}
For both the tunneling processes starting in an $\mathcal{S}$-site and those starting in a $\mathcal{P}$-site, there are two types of fourth-order tunneling processes, those which  tunnel out and back along the same path (non-loop processes) and those which makes a loop. In the non-loop process, three different lattice sites are involved in the tunnelling process, while for loop processes there are four different lattice sites involved. In eq.~(\ref{contribution2}), the subscripts $(i,k)$ refer to $\mathcal{S}$-sites and $(j,l)$ refer to $\mathcal{P}$-sites, hence when one considers a non-loop process two of these will coincide, while for loop processes they will all be different. The full expression for the effective Hamiltonian coming from fourth order transitions is the sum over all possible contributions on the above form that returns the tunnelling atom to the original lattice site by the end of the process, hence the subscripts $\alpha$ etc. on the tunneling coefficients. The $\mathcal{S}$-sites support only one type of orbital state, the $s$-orbital state, as opposed to the $\mathcal{P}$-sites which support two orbital states, the $p_{x}$ and $p_{y}$ states.  The one-level nature of the $\mathcal{S}$-site ensures that the contribution from these sites in the Mott$_1$ phase is $K_{\mathcal{O}}^{\mathcal{S}}=\hat{a}^{\dagger}_s\hat{a}_s=1$ for the origin ($\mathcal{O}$) of tunnelling process and using the bosonic commutator rules $K_{\mathcal{I}}^{\mathcal{S}}=\frac{1}{U_{ss}}\hat{a}_s\hat{a}^{\dagger}_s=\frac{2}{U_{ss}}$  for an intermediate ($\mathcal{I}$) $\mathcal{S}$-site. This allow us to simplify (\ref{contribution1}) as
\begin{equation}
\begin{array}{lll}\label{cont}
\mathcal{C}^{\mathcal{S}}&=& \frac{2}{U_{ss}}
\left[t_{\alpha_{j}}t_{\beta_{j}}\left(\hat{a}_{\alpha_{j}}\hat{K}^{\mathcal{P}}_{j}\hat{a}_{\beta_j}^{\dagger}\right)\right]
\left[t_{\alpha_{l}}^{\sigma_{3}}t_{\beta_{l}}^{\sigma_{4}*}\left(\hat{a}_{\alpha_{l}}\hat{K}^{\mathcal{P}}_{l}\hat{a}_{\beta_l}^{\dagger}\right)\right],\\ \\
\mathcal{C}^{\mathcal{P}}&=& \frac{4}{U_{ss}^{2}}\left[t_{\alpha_{j}}t_{\beta_{j}}\hat{a}^{\dagger}_{\beta_{j}}\hat{a}_{\alpha_{j}}\right]
\left[t_{\alpha_{l}}^{\sigma_{3}}t_{\beta_{l}}^{\sigma_{4}*}\left(\hat{a}_{\alpha_{l}}\hat{K}^{\mathcal{P}}_{l}\hat{a}_{\beta_l}^{\dagger}\right)\right].
\end{array}
\end{equation}
Here the subscripts $j$ and $l$ continue to refer to the two different $\mathcal{P}$-sites in the tunnelling process. In some tunneling processes this will be the same lattice site. The interaction between the $p$-orbitals $\hat{K}^{\mathcal{P}}$ depends only on the number of atoms in the lattice site. In the $\mathrm{Mott}_1$ phase there are two atoms in the intermediate lattice sites, and $\hat{K}^{\mathcal{P}}$ can be written in matrix form 
\begin{equation}\begin{array}{lll}
\hat K^\mathcal{P} & = &
\left[\begin{array}{ccc}
\hat{K}_{xx}^{xx}& \hat{K}_{xx}^{yy} & 0 \\ \\
\hat{K}_{yy}^{xx}& \hat{K}_{yy}^{yy}& 0 \\ \\
0&0& \hat{K}_{xy}^{xy}
\end{array}\right],
\end{array}
\end{equation}
where 
\begin{equation}
\begin{array}{lllll}
 \hat{K}_{\alpha\alpha}^{\alpha\alpha}=2\frac{U_{\beta\beta}}{U^2}, & & \hat{K}_{\alpha\beta}^{\alpha\beta}=\frac{1}{U_{xy}}, & & \hat{K}_{\alpha\alpha}^{\beta\beta}=-4\frac{U_{xy}}{U^{2}}\hat{a}_{\beta}^{\dagger}\hat{a}_{\beta}^{\dagger}\hat{a}_{\alpha}\hat{a}_{\alpha},
\end{array}\label{interaction_K}
\end{equation}
with $U^2=U_{xx}U_{yy}-U_{xy}^2$ \cite{fernanda}.
To express this more compactly one may define the contributing $\mathcal{P}$-sites in terms of four different processes depending on whether the transition has a loop or non-loop structure, and depending on whether the $\mathcal{P}$-site is the origin $(\mathcal{O})$ or an intermediate site $(\mathcal{I})$ in the process;
\begin{equation}\label{trans1}
\begin{array}{lll}
 \hat T_{\mathcal{O}}^{\sigma}=\sum_{\alpha\alpha'}t_{\alpha'}^{\sigma *}t_{\alpha }^{\sigma}\hat{a}_{\alpha'}^{\dagger}\hat{a}_{\alpha},&&
 \hat T_{\mathcal{I}}^{\sigma}=\sum_{\alpha\alpha}t_{\alpha'}^{\sigma *}t_{\alpha}^{\sigma}\hat{a}_{\alpha'}\hat{K}_{\mathcal{I}}^{\mathcal{P}}\hat{a}_{\alpha}^{\dagger}\\ \\
 \hat L_{\mathcal{O}}^{\sigma\sigma'}=\sum_{\alpha\alpha'}t_{\alpha'}^{\sigma *}t_{\alpha}^{\sigma'}\hat{a}_{\alpha'}^{\dagger}\hat{a}_{\alpha},&&
 \hat L_{\mathcal{I}}^{\sigma\sigma'}=\sum_{\alpha\alpha}t_{\alpha'}^{\sigma *}t_{\alpha}^{\sigma'}\hat{a}_{\alpha'}\hat{K}_{\mathcal{I}}^{\mathcal{P}}\hat{a}_{\alpha}^{\dagger}.
\end{array}
\end{equation}
For non-loop $(\hat T)$ the tunneling takes place out and back along the same path, and there is only one superscript $\sigma$, while for loops the tunneling out and back into a lattice site are along two different bonds. Then we may distinguish between four different tunneling processes. Two which start in an $\mathcal{P}$ site, where one of them will be loop $(\hat\mathcal{C}^{\mathcal{P}}_{L})$  and one will be a non-loop $(\hat\mathcal{C}^{\mathcal{P}}_{T})$, and two which start in an $\mathcal{S}$-site, with similar structure. Then (\ref{cont}) may be expressed in terms of (\ref{trans1})
\begin{equation}
\begin{array}{lll}
 \hat \mathcal{C}^{\mathcal{S}}_{T}=\frac{2}{U_{ss}}\hat T_{\mathcal{I}_{j}}^{\sigma}\hat T_{\mathcal{I}_{l}}^{\sigma},&&
 \hat \mathcal{C}^{\mathcal{S}}_{L}=\frac{4}{U_{ss}^2}\hat L_{\mathcal{I}_{j}}^{\sigma\sigma'}\hat L_{\mathcal{I}_{l}}^{\sigma},\\ \\
 \hat \mathcal{C}^{\mathcal{P}}_{T}=\frac{2}{U_{ss}}\hat T_{\mathcal{O}_{j}}^{\sigma}\hat T_{\mathcal{I}_{l}}^{\sigma},&&
 \hat \mathcal{C}^{\mathcal{P}}_{L}=\frac{4}{U_{ss}^2}\hat L_{\mathcal{O}_{j}}^{\sigma\sigma'}\hat L_{\mathcal{I}_{l}}^{\sigma}.
\end{array}
\end{equation}
Summing over neighbouring pairs of $\mathcal{P}$-sites (j,l) connected via a loop or a non-loop transition then leads to an effective Hamiltonian of the form~(\ref{HamMott}). Using (\ref{interaction_K}) the  coupling constants may be evaluated. In the isotropic lattice  $h^{z}=0$ but $ J_1^{X}, J_1^{Y}$ and $J_1^{Z}$ are given by
\begin{equation}
\begin{array}{l}
 \displaystyle{J_1^{X}=4t^4\left[\frac{V_-}{U_{ss}^2}-\frac{V_-^2}{U_{ss}}\right]},\\ \\
 \displaystyle{J_1^{Y}=4t^4\left[\frac{V_+}{U_{ss}^2}+\frac{V_+^2}{U_{ss}}\right]},\\ \\
 \displaystyle{J_1^{Z}=2t^4\left[\frac{4V_z}{U_{ss}^2}+\frac{V_{z}^2}{U_{ss}}\right]},
\end{array}
\end{equation}
with $V_\pm= 120 (U_{xy}\pm1)/U^2$ and $V_{z}=4\tilde U/U^2+1/(2U_{xy})$, and where $U_{xx}= U_{yy}= U$ and $\tilde U = U^2-U_{xy}^2$. Remember further that $J_1^\alpha=J_2^\alpha$. Analytic expressions for these can be found in the harmonic approximation where $U_{xx}=U_{yy}=3U_{xy}$ leading to coupling coefficients 
\begin{equation}
\begin{array}{l}
 \displaystyle{J^{X}_{1}= \frac{4\cdot 13 t^4}{U_{xy}U_{ss}}\left[\frac{1}{U_{ss}}-\frac{13}{U_{xy}}\right]\approx \frac{52t^4}{U_{ss}U_{xy}}\left(\frac{1}{U_{ss}}-\frac{13}{U_{xy}}\right)},\\ \\
 \displaystyle{J^{Y}_{1}= \frac{4\cdot 41t^4}{3U_{xy}U_{ss}}\left[\frac{1}{U_{ss}}+\frac{41}{3U_{xy}}\right] \approx \frac{56t^4}{U_{ss}U_{xy}}\left(\frac{1}{U_{ss}}+\frac{14}{U_{xy}}\right)},\\ \\
 \displaystyle{J^{Z}_{1}= \frac{16\sqrt{2}+9}{9}\frac{t^4}{U_{xx}U_{yy}}\left[\frac{4}{U_{ss}}+\frac{16\sqrt{2}+9}{18U_{xy}}\right]\approx \frac{16t^4}{U_{ss}U_{xy}}\left(\frac{1}{U_{ss}}+\frac{1}{U_{xy}}\right)}.
\end{array}
\end{equation}
To find the analytical expressions for the tunneling $t$ in the harmonic approximation one would need to give the explicit form of the optical lattice. Nevertheless, given the Wannier functions
\begin{equation}
\begin{array}{l}
\displaystyle{w_{s\mathbf{j}}({\bf x})=\left(\frac{1}{\pi\sigma^4}\right)^{1/2}\exp\left(-\frac{(x-x_{j_x})^2}{2\sigma^2}-\frac{(y-y_{j_y})^2}{2\sigma^2}\right)},\\ \\
\displaystyle{w_{x\mathbf{j}}({\bf x})=\left(\frac{2}{\pi\sigma^4}\right)^{1/2}(x-x_{j_x})\exp\left(-\frac{(x-x_{j_x})^2}{2\sigma^2}-\frac{(y-y_{j_y})^2}{2\sigma^2}\right)},
\end{array}
\end{equation}
in the harmonic approximation we find the interaction strengths expressed in terms of the width $\sigma$ as
\begin{equation}
U_{ss}=U_0\frac{1}{2\pi\sigma^6},\hspace{1cm}U_{xx}=U_{yy}=3U_{xy}=U_0\frac{3}{8\pi\sigma^2}.
\end{equation}


\begin{thebibliography}{999}
\bibitem{com2} It is enough if either of the two exchange interaction terms is anti-ferromagnetic to cause frustration, but how our model is constructed the two terms have equal couplings.
\end{thebibliography}
\end{document}